\documentclass[12pt,letterpaper]{JHEP3}

\usepackage{amssymb}
\usepackage{amsmath,amsfonts}
\usepackage{graphicx}

\def\KeyWord#1{$\backslash$\IfColor{$\!\!$\textRed{#1}\textBlack}{#1}$\!\!$}

\providecommand{\slashD}{D\mspace{-12mu}\slash\mspace{3mu}}

\def\beq{\begin{eqnarray}}
\def\eeq{\end{eqnarray}}
\def\bea{\begin{eqnarray*}}
\def\eea{\end{eqnarray*}}

\def\centeron#1#2{{\setbox0=\hbox{#1}\setbox1=\hbox{#2}\ifdim
\wd1>\wd0\kern.5\wd1\kern-.5\wd0\fi
\copy0\kern-.5\wd0\kern-.5\wd1\copy1\ifdim\wd0>\wd1
\kern.5\wd0\kern-.5\wd1\fi}}
\def\ltap{\;\centeron{\raise.35ex\hbox{$<$}}{\lower.65ex\hbox{$\sim$}}\;}
\def\gtap{\;\centeron{\raise.35ex\hbox{$>$}}{\lower.65ex\hbox{$\sim$}}\;}

\newcommand{\newc}{\newcommand}
\newc{\qbar}{{\overline q}}
\newc{\Kahler}{K\"ahler }
\newc{\deltaGS}{\delta_{\rm GS}}

\title{Infrared Divergences in dS/CFT}

\author{T. Banks \\
Department of Physics and Astronomy - NHETC\\
Rutgers University, Piscataway, NJ 08540\\
and\\
Department of Physics, SCIPP\\
University of California, Santa Cruz, CA 95064\\
E-mail: \email{banks@scipp.ucsc.edu}}

\author{L. Mannelli \\
Department of Physics, SCIPP\\
University of California, Santa Cruz, CA 95064\\
and\\
Kavli Institute for Theoretical Physics\\
University of California\\
Santa Barbara, CA 93106\\
E-mail: \email{lorenzo@scipp.ucsc.edu}}

\author{W. Fischler, \\
Department of Physics\\
University of Texas, Austin, TX 78712\\
E-mail: \email{fischler@utexas.edu}}

\abstract{ dS/CFT gives a perturbatively gauge invariant definition of
particle masses in de Sitter (dS) space.  We show, in a toy model in which
the graviton is replaced with a minimally coupled massless scalar field,
that loop corrections to these masses are infrared (IR) divergent.   We
argue that this implies anomalous dependence of masses on the cosmological
constant, in a true theory of quantum gravity.  This is in accord with the
hypothesis of Cosmological SUSY Breaking (CSB).}


\preprint{\hepth{} \\SCIPP 05/31\\ UTTG-02-05}

\begin{document}



\section{\textbf{Introduction}}

The hypothesis of Cosmological Supersymmetry Breaking (CSB) is based on the
idea \cite{Banks:2000fe} \cite{Banks:2001zj} \cite{Fischler:2000} that
quantum theories of stable, asymptotically de Sitter (AsdS) space-times
exist and have a finite number of physical states. The (positive)
cosmological constant, $\Lambda$,is an input parameter, which controls the
number of states. The limit of vanishing $\Lambda$ is a super-Poincare
invariant theory, but SUSY is broken for finite $\Lambda$: the operator
which converges to the Poincare Hamiltonian $P_0$, does not commute with the
SUSY charges.

Classical SUGRA supports such a picture, but suggests a relation between the
gravitino mass and the c.c.: $m_{3/2} \sim \Lambda^{1/2} / M_P$. CSB is the
proposal that the exponent $1/2$ in this relation is replaced by $1/4$ in
the quantum theory. Given the interpretation of $\Lambda$ as a parameter
controlling the number of states, this is a critical exponent, and it is
plausible that it has fluctuation corrections. Indeed, low energy effective
field theory cannot calculate the real relation between the gravitino mass
and the c.c., since the c.c. is a relevant parameter and one must introduce
a counterterm for it. The exponent above is just the ``natural" relation of
classical SUGRA, without fine tuning of the constant in the superpotential.
If we accept such fine tuning, we can get any relation we want between $%
m_{3/2}$ and $\Lambda$ in effective field theory.

However, the necessity of canceling an infinite c.c. appears to be a short
distance problem in effective field theory, and as such, does not seem to
depend on the value of the c.c. . As a consequence, there has been
considerable skepticism about CSB.

In \cite{Banks:2002wj}, one of the authors presented an argument for the
exponent $1/4$, based on crude approximations to the dynamics of the
cosmological horizon in the static observer gauge for dS space. It was clear
that from the static observer's point of view, the enhanced exponent is an
IR effect. However, since the argument relied on conjectures about the
horizon dynamics, it has not convinced anyone. Skeptical observers want to
understand where effective field theory reasoning breaks down. The work of
\cite{Banks:2002wr} provided an important clue. In the static gauge most of
the states in a quantum theory of dS space live on the horizon of the static
observer. Local field theory can describe only a negligible fraction of the
entropy. On the contrary, it was argued in \cite{Banks:2002wr} that in
global coordinates, the entire Hilbert space may be well described by field
theory. The contradiction between a finite number of states and the field
theoretic description can be viewed as an IR cutoff, which restricts the
global time coordinate to an interval of order $|t| \leq {\frac{R}{6}}
\mathrm{ln} \left(R M_P\right)$ around the time symmetric point. The field
theory also has a UV cutoff at a scale $M_c \sim \left({\frac{M_P }{R}}%
\right)^{1/2}$. This description is inappropriate for states containing
black holes whose size scales like $R$, but there is a basis of field
theoretic states in global coordinates, which may span the Hilbert space.

A simple way to restate this conclusion is to invoke the fact that the
global description of dS space in field theory does not seem to break down
until we contemplate introducing black holes on early time initial data
slices, whose entropy exceeds that of the space-time. The combined UV and IR
cutoffs prevent us from introducing such objects, and describes a cutoff
field theory with a finite number of states. The field theory description of
many of these states breaks down near the time-symmetric point, but near the
upper and lower limits of $t$, it is a good approximation to their
properties.

We have thus set up a framework in which IR divergences in a field theoretic
treatment of dS space can be thought of as introducing non-classical
dependence on the c.c. . It has often been argued that perturbative quantum
gravity expanded around dS space is fraught with IR divergences. These
claims have been less than convincing, because no-one had identified gauge
invariant observables with which to check the physical meaning of the
logarithmically growing graviton propagator. This problem is solved by
dS/CFT \cite{Witten:2001kn}\cite{Strominger:2001pn}\cite{Maldacena:2002vr}.
In particular, the mass of a field in dS space is given a gauge invariant
meaning: it is related to the dimension of a conformal field on the boundary.

The plan of this paper is as follows: in the next section we review dS/CFT,
in the Wheeler-DeWitt formalism proposed by Maldacena. This allows
perturbative calculations to be performed in a straightforward manner,
apparently troubled only by conventional UV divergences. In section 3 we
perform one loop calculations of boundary dimensions in a variety of
non-gravitational field theories. We find that when the theory contains a
massless, minimally coupled scalar field with soft couplings, the dimensions
are infected with IR logarithms. In the conclusions, we discuss the
difficulties attendant on an extension of these calculations to perturbative
quantum gravity.

\section{Review of dS/CFT}

In his talk at Strings 2001 in Mumbai \cite{Witten:2001kn}, Witten proposed
a sort of scattering theory for de Sitter space. The fundamental object was
the path integral with fixed boundary conditions on $\mathcal{I}_{\pm}$. It
was implicitly assumed that, as in asymptotically flat and Anti-deSitter
spaces, a field theoretic approximation became exact near the boundaries of
space-time. This assumption is open to criticism. It is likely that generic
boundary conditions on fields on $\mathcal{I}_-$ will lead to Big Crunch
space-times, rather than space-times which are future asymptotically dS.
However, this criticism does not apply to perturbation theory, where the
boundary conditions are infinitesimal perturbations of those corresponding
to the dS vacuum. Witten's prescription provides a perturbative definition
of amplitudes in dS quantum gravity, which are invariant under
diffeomorphisms that approach the identity near $\mathcal{I}_{\pm}$.

Somewhat later, Strominger proposed \cite{Strominger:2001pn} that suitably
defined boundary amplitudes should be the correlation functions of a
Euclidean conformal field theory (CFT). An apparent difference with Witten's
proposal is the role of conformally covariant, rather than invariant
amplitudes in dS/CFT. However, Maldacena \cite{maldapriv} has emphasized
that the operator dimensions, OPE coefficients and the like, of dS/CFT, are
gauge invariant observables in the sense of Witten.

The boundary correlation functions defined by Strominger should certainly be
conformally invariant, but it is not clear that they should obey the axioms
of field theory. Analogous arguments would lead us to believe that the
holographic dual of linear dilaton backgrounds \cite{Aharony:1999ti} was a
Lorentz invariant field theory. The calculations of Peet and Polchinski \cite%
{Peet:1998wn} show that it is not. In the dS/CFT case, the form of the two
point function follows from conformal invariance alone, and does not give us
enough of a clue to the nature of the holographic dual. As believers in the
proposition that quantum dS space has only a finite number of states, the
present authors are inclined to disbelieve that a CFT will be the exact
description of the quantum theory.

For our present purposes, all of these issues of principle are somewhat
beside the point. We want a definition of correlation functions on $\mathcal{%
I}_{\pm}$ which is perturbatively well defined and gauge invariant.
Furthermore, we will be interested only in two point functions, and will not
have to address the question of whether higher order correlators obey the
axioms of CFT. We have found that the dS/CFT prescription advocated by
Maldacena \cite{Maldacena:2002vr} is the most appropriate for our purposes.
Maldacena observes that the Euclidean path integral on a space with the
topology of a hemisphere defines a ``wave function of the universe" which is
a functional of fields on the boundary of the hemisphere. In leading
semiclassical approximation, the geometry is the section of the round sphere
metric
\begin{equation*}
{ds^2 = d\theta^2 + sin^2 (\theta ) d\Omega^2 }
\end{equation*}
with $0 \leq \theta \leq \theta_0$. Maldacena defines boundary correlators
as the expansion coefficients of the logarithm of the wave function of the
universe for fixed $\theta_0$. The analytic extrapolation $\theta_0
\rightarrow {\frac{\pi}{2}} + i t$, $ t\rightarrow\infty$ defines
correlation functions on $\mathcal{I}_+$. If the limiting correlation
functions exist, they should be covariant under the conformal group of the
sphere. In particular, if we work in planar coordinates on the upper
triangle of the dS Penrose diagram
\begin{equation*}
ds^{2}=\frac{1}{\eta ^{2}}\left( -d\eta ^{2}+d\mathbf{x}^{2}\right)
\end{equation*}
($\mathcal{I}_+$is at $\eta = 0$) then the boundary two point function
should have the form $|\mathbf{x}|^{-\Delta} $. For a free scalar field of
mass $m^2$ this is indeed true, and the relation between mass and dimension
is given by
\begin{equation*}
\Delta _{\pm }=a=\frac{1}{2}\left( d-1\pm \sqrt{(d-1)^{2}-4m^{2}R^{2}}\right)
\end{equation*}
This is an analytic continuation (in the c.c.) of analogous formulas in
AdS/CFT. Indeed, Maldacena's proposal for the correlation functions is the
direct analog of the calculation of Euclidean correlation functions in
AdS/CFT.

The purpose of the present paper is to compute one loop corrections to $%
\Delta_{\pm}$ in simple field theory models. We will see that when the
theory has a massless, minimally coupled scalar with soft couplings, these
corrections are IR divergent.

\section{Review of QFT in dS space}

In this section we will introduce the principal formulae of QFT in\textit{d-}%
dimensional de Sitter (dS$^{d}$) space, and fix our notation .

For a more complete discussion we refer to the excellent review paper \cite%
{Spradlin:2001pw}.

\subsection{Coordinate Systems}

\textit{d-}dimensional de Sitter dS$^{d}$ can be realized as the manifold,
embedded in $d+1$ dimensional Minkowski $\mathcal{M}^{d,1}$ space, defined
by the equation%
\begin{equation}
-X_{0}^{2}+X_{1}^{2}+\cdots X_{d}^{2}=R^{2}  \label{Eq for
Hyperboloid}
\end{equation}%
where $R$ is the de Sitter radius.

The de Sitter metric is the standard metric induced by immersion in $%
\mathcal{M}^{d,1}$ with the usual flat metric. The isometry group of dS$^{d}$
is $O(d,1)$ in fact this leave invariant both the hyperboloid defined by the
equation (\ref{Eq for Hyperboloid}) and the flat metric of $\mathcal{M}%
^{d,1} $.

For the most part, we will use planar coordinates%
\begin{eqnarray}
X^{0} &=&\sinh t-\frac{1}{2}x_{i}x_{i}e^{-t}  \notag \\
X^{i} &=&x^{i}e^{-t}  \label{coordinates: planar} \\
X^{d} &=&\cosh t-\frac{1}{2}x_{i}x_{i}e^{-t}  \notag
\end{eqnarray}%
with $i=1,\ldots ,d$ the metric take the form%
\begin{equation*}
ds^{2}=-dt^{2}+e^{-2t}dx_{i}dx_{i}  \label{metric: planar}
\end{equation*}%
In these coordinates the spatial sections have flat \textit{Euclidean}
metric.

It is useful to introduce conformal coordinates too, defined by the
transformation%
\begin{equation*}
\eta =e^{t}
\end{equation*}%
The metric is conformally flat and takes the form%
\begin{equation*}
ds^{2}=\frac{1}{\eta ^{2}}\left( -d\eta ^{2}+dx_{i}dx_{i}\right)
\label{metric: Lorentz. conf. flat}
\end{equation*}%
with $i=1,\ldots ,d$. In the following, unless otherwise stated, we will
consider the \textit{Euclidean} section of dS$^{d}$ defined by the
analytical continuation%
\begin{equation}
\eta \rightarrow ix_{0}  \label{analitical cont.}
\end{equation}%
after the transformation (\ref{analitical cont.}) the metric become%
\begin{equation}
ds^{2}=-\frac{1}{x_{0}^{2}}\left( dx_{0}^{2}+dx_{i}dx_{i}\right)
\label{metric: Euclid. conf. flat}
\end{equation}%
in these coordinates the boundary of dS$^{d}$ $\Sigma $ is given by the
submanifold $x_{0}=\epsilon $ where $\epsilon \rightarrow 0$.

\subsection{Geodesic Distance\label{Sect.:Geodesic Distance}}

The geodesic distance between two points $x$ and $x^{\prime }$ is
\begin{equation*}
\mu (x,x^{\prime })=\int_{0}^{1}\left[ g_{ab}\dot{x}^{a}(\lambda )\dot{x}%
^{b}(\lambda )\right] ^{\frac{1}{2}}d\lambda
,~~x^{a}(0)=x,~~x^{a}(1)=x^{\prime }
\end{equation*}
In the following we will often use the new variable%
\begin{equation*}
z=\cos ^{2}\left( \frac{\mu }{2R}\right)  \label{geodes. dist.:z}
\end{equation*}%
It is possible to show that%
\begin{equation*}
\cos \left( \frac{\mu (x,x^{\prime })}{R}\right) =\frac{\eta
_{ab}X^{a}(x)X^{b}(x^{\prime })}{R^{2}}
\end{equation*}%
with $X^{a}(x),~X^{b}(x^{\prime })\in \mathcal{M}^{d,1}$ embedding
coordinates and $\eta _{ab}=$diag$(-1,1,\ldots ,1)$.

Consequently we have%
\begin{eqnarray*}
z &=&\cos ^{2}\left( \frac{\mu }{2R}\right) \\
&=&\frac{1}{2}\left( 1+\cos (\frac{\mu }{R})\right) \\
&=&\frac{1}{2}\left( 1+\frac{\eta _{ab}X^{a}(x)X^{b}(x^{\prime })}{R^{2}}%
\right)
\end{eqnarray*}%
In the \textit{Euclidean} conformally flat coordinates (\ref{metric: Euclid.
conf. flat}) we have%
\begin{equation*}
z=-\frac{(x_{0}-y_{0})^{2}+\left( \bar{x}-\bar{y}\right) ^{2}}{x_{0}y_{0}}%
=-2+\frac{x_{0}^{2}+y_{0}{}^{2}+(\bar{x}-\bar{y})^{2}}{x_{0}y_{0}}
\label{z: euclid. conf. flat coord.}
\end{equation*}

\subsection{The Cut-off Prescription}

Maldacena's prescription defines the boundary correlators by analytic
continuation in global time. We have proposed that these formulae should be
cut off at a fixed global time $T$. IR divergences will appear as divergent
behavior at large $T$. It is most convenient to do calculations in conformal
coordinates. Thus we have to understand the effect of a global time cut-off
in conformal coordinates.

The relation between the two coordinate systems is most simply understood by
writing the embedding coordinates in terms of conformal coordinates. The
slices of fixed embedding time and global time coincide:

\begin{figure}[tbp]
\begin{minipage}[t]{7cm}
\begin{center}
\includegraphics[width=7.0cm,clip]{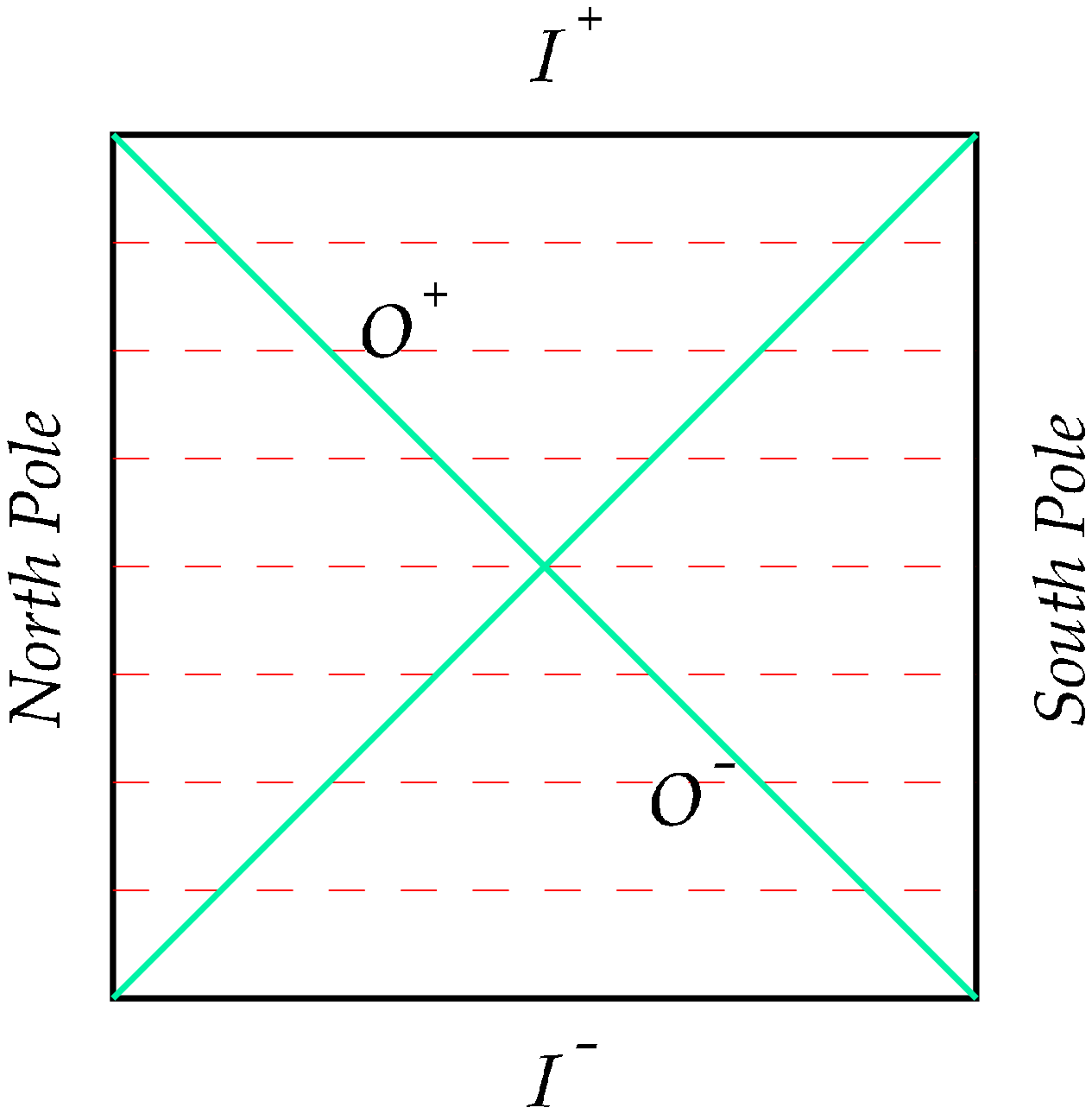}
\caption[Short caption for figure 2]{ \textbf{Global coordinates.} {\em
Foliation of dS with compact spatial sections (spheres).}}
\label{PenroseGlobal}
\end{center}
\end{minipage}
\hfill
\begin{minipage}[t]{7cm}
\begin{center}
\includegraphics[width=7.0cm,clip]{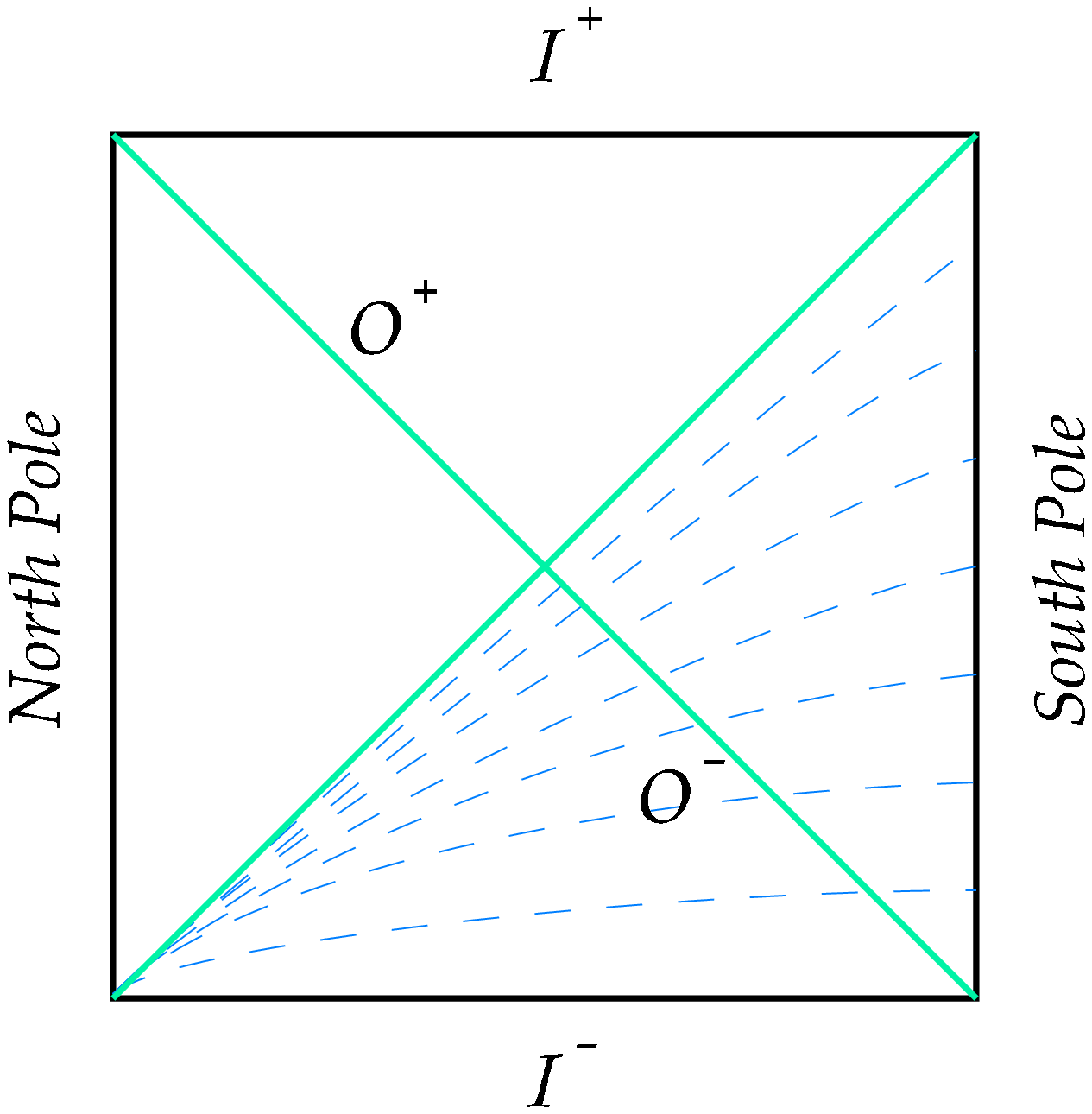}
\caption[Short caption for figure 1]{\textbf{Flat coordinates.} {\em
Foliation of dS with flat spatial sections.}} \label{PenroseFlat}
\end{center}
\end{minipage}
\end{figure}

\begin{equation*}
X^{0}=\frac{R}{2}\left( \frac{x^{0}}{R}-\frac{R}{x^{0}}\right) -\frac{%
\mathbf{x}^{2}}{2x^{0}}  \label{embedconformal}
\end{equation*}%
At $X^0 = T$, see Fig. \ref{PenroseGlobal} and Fig. \ref{PenroseFlat}. This
relation implies a maximal value of $|\mathbf{x}|$ for fixed $x^0$, as well
as a maximal value of $x^0$ (which runs between $-\infty$ and $0$ in the
conformal coordinate patch). The relation is
\begin{equation*}
x_{\max }^{2}=-2x^{0}\left( T-x^{0}+\frac{R^{2}}{x^{0}}\right)
\end{equation*}%
The maximal value of $x^0$ is the point at which $x_{max} = 0$.

\begin{equation*}
x_{max}^0 \approx - {\frac{R^2 }{T}} \ \ \ T \gg R
\end{equation*}
The maximal geodesic distance between two points on a give $x^0$ slice is ${%
\frac{x_{max} }{x_{max}^0}}$. The slice on which this distance is maximal is
given by $x^0_* = - {\frac{2 R^2 }{T}}$. The geodesic distance on this slice
is $o(T)$, while the maximum coordinate distance is $o(R)$. IR divergences
will come predominantly from slices near this maximal slice.

Dirichlet boundary conditions on the $X^0 = T$ surface become \textit{spatial%
} Dirichlet boundary conditions on the spatial slices of conformal
coordinates. On most of the slice of maximal geodesic size, the Dirichlet
propagator will coincide with the usual Euclidean propagator defined by
analytic continuation from the entire sphere. Thus, the boundary conditions
will not affect the IR divergences.

\subsection{Wave Function of the Universe\label{Sect.: Wave Function of the
Universe}}

We are looking for a gauge invariant definition of the IR\ renormalization
of the particle mass. The Wave Function of the Universe (WFU) will provide
us with such a definition.

The WFU $\Psi \lbrack h_{ij},\phi _{0}]$ was first introduced by Hartle and
Hawking in \cite{Hartle:1983ai}. If $I[g,\phi ]$ is the Euclidean action for
gravity and a set of fields indicated by $\phi $, the \textit{Euclidean} WFU
is defined as the path integral%
\begin{equation}
\Psi \lbrack h_{ij},\phi _{0}]=\int_{C}[dg][d\phi ]e^{-I[g,\phi ]}~
\label{Wave Function Universe: Eucl.}
\end{equation}%
over a class $C$ of space-times with a compact space-like boundary $\Sigma $
on which the induced metric is $h_{ij}$ and over the field configurations $%
\phi $ with boundary value $\phi _{0}$. The boundary $\Sigma $ has only one
connected component.

In the case $\Lambda >0$ we imagine a semiclassical expansion of the
integral over Riemannian spaces with the topology of a hemisphere, expanded
around the metric on the portion of the round sphere below polar angle $%
\theta_0$. We then analytically continue to the future half of Lorentzian dS
space. This prescription corresponds to the choice of Euclidean vacuum in de
Sitter space.

Given the WFU we can define the "boundary two-point function" in the limit
where the boundary is taken to $\mathcal{I^+}$%
\begin{equation*}
\frac{\delta \Psi \lbrack h_{ij},\phi _{0}]}{\delta \phi _{0}(\bar{x})\delta
\phi _{0}(\bar{y})}
\end{equation*}%
Once we expand around dS$^{d}$ we find
\begin{equation}
\frac{\delta \Psi \lbrack h_{ij},\phi _{0}]}{\delta \phi _{0}(\bar{x})\delta
\phi _{0}(\bar{y})}=C_+\frac{1}{(\bar{x}-\bar{y})^{2\Delta_+ }} + C_- \frac{1%
}{(\bar{x}-\bar{y})^{2\Delta_- }}  \label{Bdry 2-point fnct:
dS}
\end{equation}%
, where $C_{\pm}$ are constants This form is dictated by conformal
invariance. If $\lambda $ and $m$ are the coupling and the mass of the field
$\phi $, in the classical Lagrangian, then $\Delta $ will be a function of $%
\lambda $ and $m$ and will provide a gauge invariant definition of the
renormalized mass.

The Eq. (\ref{Bdry 2-point fnct: dS}) is the analogue of the boundary
correlators defined in the AdS/CFT correspondence%
\begin{equation*}
Z[\phi _{0}]=\left\langle e^{\int d^{4}x\phi _{0}(x)O(x)}\right\rangle
_{CFT}~\ ,~\phi (x_{0}=\epsilon )\sim \phi _{0}
\end{equation*}%
\begin{equation*}
\langle 0|O(\bar{x})O(\bar{y})|0\rangle =\frac{\delta Z}{\delta \phi _{0}(%
\bar{x})\delta \phi _{0}(\bar{y})}=\tilde{C}\frac{1}{(\bar{x}-\bar{y})^{2%
\tilde{\Delta}}}
\end{equation*}

There are however, important differences between the two cases. They stem
from the fact that the Euclidean section of dS space is a spherical cap and
has a conventional Dirichlet problem, different from the singular Dirichlet
boundary conditions on the boundary of Euclidean AdS space. There are no
large volume divergences in the Euclidean calculation. They appear only
after extrapolation to infinite Lorentzian time. As a consequence, the
divergent behavior comes as a combination of both powers $\Delta_{\pm}$. For
fields corresponding to the principal series of dS representation theory,
the real parts of $\Delta_{\pm}$ are equal.

The prescription to extract boundary two-point function in dS$^{d}$ given by
(\ref{Bdry 2-point fnct: dS}) was first pointed out by Maldacena in \cite%
{Maldacena:2002vr} and it is, as explained in this paper, different from the
prescription used by Strominger and collaborators in \cite{Strominger:2001pn}%
, \cite{Bousso:2001mw}.

\subsection{Representations of the dS$^{d}$ Group\label{Representions of dS
Group}}

The scalar representation of the de Sitter group $SO(1,d)$ are classified
according to the mass $m$ in the following series, see \cite{Tolley:2001gg},
\cite{Gazeau:1999mi}: the principal series%
\begin{equation*}
m^{2}\geqslant \left( \frac{d-1}{2R_{dS}}\right)  \label{principal
series}
\end{equation*}%
the complementary series
\begin{equation*}
0<m^{2}<\left( \frac{d-1}{2R_{dS}}\right)  \label{complementary
series}
\end{equation*}%
and the discrete series, whose only case of physical interest is $m^{2}=0$.

Under a Wigner-In\"{o}n\"{u} contraction to the Poincare group, only the
representations of the principal series contract to representation of the
Poincare\ Group.

Lowe and G\"{u}ijosa \cite{Guijosa:2003ze} and Lowe \cite{Lowe:2004nw} use
the principal series to construct the dS/CFT correspondence. They stress the
fact that when one replaces the dS isometry group with a \textit{q}-deformed
version, the unitary principal representation deform to a finite dimensional
unitary representation of the quantum group\footnote{%
The idea that a q-deformed version of the dS group might have finite
dimensional unitary representations, resolving the contradiction between dS
invariance and a finite number of states, was pointed out to one of the
authors (TB) by A. Rajaraman in the fall of 1999. There seemed to be a
problem with this idea, because the dS group has no highest weight unitary
representations, but Lowe and G\"{u}ijosa made the crucial observation that
the cyclic representations of the quantum group (which are not highest
weight) converged to the principal unitary series.}.

The massive scalar particles in our formulae will always correspond to the
principle series representations, so that the boundary dimensions all have
the same real part. We will also use a massless, minimally coupled scalar,
which is our toy model of the graviton.

\section{Scalar Green Functions}

In the next few subsections we will derive the scalar Green Functions
relevant for our computations and their asymptotic behavior. As explained
in the section on the cut-off procedure, we will not impose Dirichlet
boundary conditions on the bulk propagators. The IR divergences, which are
our principal concern, are not affected by the boundary conditions on the
bulk propagator. For a more detailed discussion of dS Green functions, see
for example \cite{Allen:1985wd}, \cite{Kirsten:1993ug}.

\subsection{Maximally Symmetric Bitensors}

The relevant geometric objects in maximally symmetric spaces, like dS, are
the geodesic distance $\mu (x,x^{\prime })$ between two points $x$ and $%
x^{\prime }$, the unit tangent vectors $n_{\sigma }(x,x^{\prime })$ and $%
n_{\sigma ^{\prime }}(x,x^{\prime })$ to the geodesic at $x$ and at $%
x^{\prime }$, the vector parallel propagator $g^{\mu }{}_{\nu ^{\prime
}}(x,x^{\prime })$ and the spinor parallel propagator $\Lambda ^{\alpha
}{}_{\beta ^{\prime }}(x,x^{\prime })$.

The geodesic distance is by definition the distance along the geodesic $%
x^{a}(\lambda )$ connecting $x$ and $x^{\prime }$%
\begin{equation*}
\mu (x,x^{\prime })=\int_{0}^{1}\left[ g_{ab}\dot{x}^{a}(\lambda )\dot{x}%
^{b}(\lambda )\right] ^{\frac{1}{2}}d\lambda
,~~x^{a}(0)=x,~~x^{a}(1)=x^{\prime }
\end{equation*}%
The vectors $n_{\sigma },n_{\sigma ^{\prime }}$ are defined by
\begin{equation*}
n_{\sigma }=\nabla _{\sigma }\mu (x,x^{\prime })\qquad \mathrm{and}\qquad
n_{\sigma ^{\prime }}=\nabla _{\sigma ^{\prime }}\mu (x,x^{\prime })\
\end{equation*}%
where $\nabla _{\sigma }$ is the covariant derivative. We note that
\begin{equation*}
n_{\sigma }=-g_{\sigma }{}^{\rho ^{\prime }}n_{\rho ^{\prime }}\
\end{equation*}%
The vector and spinor parallel propagators are defined by
\begin{eqnarray}
V^{\mu }(x) &=&g^{\mu }{}_{\nu ^{\prime }}(x,x^{\prime })V^{\nu ^{\prime
}}(x^{\prime })\   \label{parallel propag.: vector} \\
\psi ^{\alpha }(x) &=&\Lambda ^{\alpha }{}_{\beta ^{\prime }}(x,x^{\prime
})\psi ^{\beta ^{\prime }}(x^{\prime })  \label{parallel propag.: spinor}
\end{eqnarray}%
for every parallel-transported vector $V^{\mu }(x)$ and spinor $\psi
^{\alpha }(x)$, respectively.

Tensors that depend on two points $x$ and $x^{\prime }$ on the manifold are
called \textit{bitensor. }We will say that a bitensor is \textit{maximally
symmetric} if is invariant under any isometry of the manifold. It can be
proved that any maximally symmetric bitensor can be expressed as a sum of
products of $g^{\mu }{}_{\nu ^{\prime }},~g_{\mu \nu },~g_{\mu ^{\prime }\nu
^{\prime }}{},~\mu ,~n_{\sigma }$ and $n_{\sigma ^{\prime }}$. Furthermore
the coefficients of these terms are functions only of the geodesic distance $%
\mu (x,x^{\prime })$.

The covariant derivatives of the above bitensors are given by
\begin{eqnarray}
\nabla _{\mu }n_{\nu } &=&A\,(g_{\mu \nu }-n_{\mu }n_{\nu })\   \notag \\
\nabla _{\mu ^{\prime }}n_{\nu } &=&C\,(g_{\mu ^{\prime }\nu }+n_{\mu
^{\prime }}n_{\nu })\   \notag \\
\nabla _{\mu }g_{\nu \rho ^{\prime }} &=&-(A+C)\,(g_{\mu \nu }n_{\rho
^{\prime }}+g_{\mu \rho ^{\prime }}n_{\nu })\
\label{foundam. bitensors: derivatives} \\
\nabla _{\mu }\Lambda ^{\alpha }{}_{\beta ^{\prime }} &=&\frac{1}{2}%
(A+C)\,[\,(\Gamma _{\mu }\Gamma ^{\nu }n_{\nu }-n_{\mu })\,\Lambda
]^{\,\alpha }{}_{\beta ^{\prime }}\   \notag \\
\nabla _{\mu ^{\prime }}\Lambda ^{\alpha }{}_{\beta ^{\prime }} &=&-\frac{1}{%
2}(A+C)\,[\,(\Gamma _{\mu ^{\prime }}\Gamma ^{\nu ^{\prime }}n_{\nu ^{\prime
}}-n_{\mu ^{\prime }})\,\Lambda ]^{\,\alpha }{}_{\beta ^{\prime }}\   \notag
\end{eqnarray}%
where $A$ and $C$ are the following functions of the geodesic distance:
\begin{eqnarray}
&&\mathrm{for\,\,\,\,\mathbf{R}^{d}:}\qquad A(\mu )=\frac{1}{\mu }\,\qquad
C(\mu )=-\frac{1}{\mu }\,  \notag \\
&&\mathrm{for\,\,\,\,dS~and~AdS:}\qquad A(\mu )=\frac{1}{R}\cot \frac{\mu }{R%
}\,\qquad C(\mu )=-\frac{1}{R\sin \left( \frac{\mu }{R}\right) }\
\label{table: A, C for R, dS and AdS}
\end{eqnarray}%
The radius $R$ is real for dS$^{d}$ and it is $R=i\tilde{R}$ with $\tilde{R}$
real for AdS$^{d}$. The covariant gamma matrices satisfy the usual relation $%
\{\Gamma ^{\mu },\Gamma ^{\nu }\}=2g^{\mu \nu }$.

\subsection{Bulk Two-Point Function\label{Scalar Two-Point Function}}

In this subsection we will evaluate the scalar two-point function%
\begin{equation*}
G(x,x^{\prime })=\langle \psi |\phi (x)\phi (x^{\prime })|\psi \rangle
\end{equation*}

We will assume that the state $|\psi \rangle $ is maximally symmetric, this
implies that for spacelike separated points $G(x,x^{\prime })$ depends only
on the geodesic distance $\mu (x,x^{\prime })$. For timelike separation the
symmetric and Feynman functions also depend only on $\mu $ but the
commutator function depend on the time ordering too. Doing an analytical
continuation from spacelike separation $\mu ^{2}>0$ to timelike separation $%
\mu ^{2}<0$, it is possible to obtain all these two-point functions.

We now derive a differential equation for $G(x,x^{\prime })$ . Applying the
Laplacian operator to $G(x,x^{\prime })$ we have%
\begin{eqnarray*}
\square G(\mu ) &=&\nabla ^{\nu }\nabla _{\nu }G(\mu )=\nabla ^{\nu
}(G^{\prime }(\mu )n_{\nu }) \\
&=&G^{\prime \prime }(\mu )n^{\nu }n_{\nu }+G^{\prime }(\mu )\nabla ^{\nu
}n_{\nu } \\
&=&G^{\prime \prime }(\mu )+(d-1)A(\mu )G^{\prime }(\mu )
\end{eqnarray*}%
where we have used the formulae (\ref{table: A, C for R, dS and AdS}) and
the notation $G^{\prime }=\frac{dG}{d\mu }$.

Using the equation of motion $(\square -m^{2})\phi =0$ we find
\begin{equation}
G^{\prime \prime }(\mu )+(d-1)A(\mu )G^{\prime }(\mu )-m^{2}G=0
\label{Eq for scalar 2-point fnct. in mu}
\end{equation}%
as long as $x\neq x^{\prime }$.

Defining the change of variable%
\begin{equation*}
z=\cos ^{2}\left( \frac{\mu }{2R}\right)
\end{equation*}%
the Eq. (\ref{Eq for scalar 2-point fnct. in mu}) for $G$ becomes%
\begin{equation}
z(1-z)\frac{d^{2}G}{dz^{2}}+[c-(a+b+1)z]\frac{dG}{dz}-abG=0
\label{Eq for scalar 2-point fnct. in z}
\end{equation}%
where we defined%
\begin{eqnarray}
a &=&\Delta _{+}=\frac{1}{2}\left( d-1+\sqrt{(d-1)^{2}-4m^{2}R^{2}}\right)
\label{Delta +} \\
b &=&\Delta _{-}=\frac{1}{2}\left( d-1-\sqrt{(d-1)^{2}-4m^{2}R^{2}}\right)
\label{Delta -} \\
c &=&\frac{1}{2}d  \label{c}
\end{eqnarray}

\subsubsection{De Sitter Space: Massive Scalar\label{SEC: dS GF}}

De Sitter space corresponds to choosing $R$ real in the Eq. (\ref{table: A,
C for R, dS and AdS}). There are two linearly independent solution $G(z)$ to
Eq. (\ref{Eq for scalar 2-point fnct. in z}). Any of the solutions of Eq. (%
\ref{Eq for scalar 2-point fnct. in z}) is associated with a particular
vacuum $|\psi \rangle $.

The Two-point function%
\begin{equation*}
G_{E}(x,x^{\prime })=\langle E|\phi (x)\phi (x^{\prime })|E\rangle
\end{equation*}%
associated with the Euclidean vacuum $|E\rangle $ Introduced in Section \ref%
{Sect.: Wave Function of the Universe} and defined as analytical
continuation from the sphere is given by%
\begin{equation}
G_{E}(x,x^{\prime })=qF(a,b;c;z)  \label{2-point fnct.: scalar, in E
vacuum}
\end{equation}%
where $F(a,b;c;z)$ is the hypergeometric function.

The two-point function in the Euclidean vacuum turns out to have the
following properties:

\begin{enumerate}
\item has only one singular point at $\mu (x,x^{\prime })=0$ and therefore
regular at $\mu (x,x^{\prime })=\pi R$

\item Has the same strength $\mu \rightarrow 0$ singularity as in flat space.
\end{enumerate}

The constant $q$ in Eq. (\ref{2-point fnct.: scalar, in E vacuum}) is
determined by the condition that as $\mu \rightarrow 0$ $G_{E}(x,x^{\prime
}) $ has to approach the flat two point function%
\begin{equation*}
G_{flat}(\mu )\sim \frac{\Gamma \left( \frac{d}{2}\right) }{2(d-2)\pi ^{%
\frac{d}{2}}}\mu ^{-d+2},~~\mu \rightarrow 0
\end{equation*}%
we find%
\begin{equation*}
q=\frac{\Gamma (a)\Gamma (b)}{\Gamma \left( \frac{d}{2}\right) 2^{d}\pi ^{%
\frac{d}{2}}}R^{-d+2}
\end{equation*}

For the computation it will be useful to derive the asymptotic expansion of $%
G(z)$ for $z\rightarrow -\infty $ that correspond to $x_{0}\rightarrow 0$ or
$y_{0}\rightarrow 0$.

The geodesic distance in conformally flat coordinate was given in Section %
\ref{Sect.:Geodesic Distance} and it is
\begin{equation*}
z=-\frac{(x_{0}-y_{0})^{2}+(\bar{x}-\bar{y})^{2}}{x_{0}y_{0}}
\end{equation*}%
we have%
\begin{equation*}
\lim_{\substack{ x_{0}\rightarrow 0  \\ y_{0}\rightarrow 0}}~z\sim -\frac{(%
\bar{x}-\bar{y})^{2}}{x_{0}y_{0}}
\end{equation*}%
so that the asymptotic expansion of $G(z)$ for $z\rightarrow -\infty $ is
\begin{equation}
\lim_{z\rightarrow \infty }G(z)\sim C_{+}\frac{1}{z^{\Delta _{+}}}+C_{-}%
\frac{1}{z^{\Delta _{-}}}=C_{+}\left( \frac{-x_{0}y_{0}}{(\bar{x}-\bar{y}%
)^{2}}\right) ^{\Delta _{+}}+C_{-}\left( \frac{-x_{0}y_{0}}{(\bar{x}-\bar{y}%
)^{2}}\right) ^{\Delta _{-}}  \label{2-point fnct.: scalar, asympt.
exp.}
\end{equation}%
with%
\begin{equation*}
C_{+}=q~\frac{\Gamma (\frac{d}{2})\Gamma (\Delta _{-}-\Delta _{+})}{\Gamma
(\Delta _{-})\Gamma (\frac{d}{2}-\Delta _{+})},~~~C_{-}=q~\frac{\Gamma (%
\frac{d}{2})\Gamma (\Delta _{+}-\Delta _{-})}{\Gamma (\Delta _{+})\Gamma (%
\frac{d}{2}-\Delta _{-})}
\end{equation*}

\subsubsection{De Sitter Space: Massless Scalar}

The two-point function for a massless minimally coupled scalar field in de
Sitter space was studied in \cite{Allen:1987tz}, \cite{Kirsten:1993ug}. They
find the following expression for the two-point function%
\begin{eqnarray}
G_{0}(z) &=&\frac{R^{2}}{192\pi ^{2}m^{2}}+\frac{R}{48\pi ^{2}}\left( \ln
(1-z)+\frac{1}{1-z}\right)  \label{2-point fnct.:
m=0, scalar, in E vacuum}
\\
&=&C_{0}\left( \ln (1-z)+\frac{1}{1-z}\right) +\tilde{C}  \notag
\end{eqnarray}

We will not need the actual values of the constants $C_{0}$ and $\tilde{C}$
in our computation.

The asymptotic expansion for $z\rightarrow -\infty $ of the massless
two-point function (\ref{2-point fnct.: m=0, scalar, in E vacuum}) is%
\begin{equation}
G_{0}(z)\sim C_{0}\left( \ln \frac{(\bar{x}-\bar{y})^{2}}{x_{0}y_{0}}+\frac{%
x_{0}y_{0}}{(\bar{x}-\bar{y})^{2}}\right)
\label{2-point fnct.: m=0,
scalar, asympt. exp.}
\end{equation}

\subsection{Bulk to Boundary Propagators: dS/AdS\label{Sect.: B-B prop.
scalar}}

The Bulk to Boundary propagator for AdS$^{d}$ were derived by Witten in \cite%
{Witten:1998qj}. They obey the equations

\begin{equation*}
(\square _{x}-m^{2})\tilde{K}(x,\bar{y})=0
\end{equation*}%
\begin{equation*}
\tilde{K}(\bar{x},x_{0};\bar{y})\rightarrow (x_{0})^{((d-1)-\Delta )}\delta
^{d}(\bar{x}-\bar{y}),~\text{for }x_{0}\rightarrow 0
\end{equation*}%
and their explicit form in the Poincare coordinates in AdS$^{d}$ is%
\begin{equation*}
\tilde{K}(\bar{x},x_{0};\bar{y})=\frac{\Gamma (\Delta )}{\pi ^{\frac{d-1}{2}%
}\Gamma \left( \Delta -\frac{d-1}{2}\right) }\left( \frac{x_{0}}{x_{0}^{2}+(%
\bar{x}-\bar{y})^{2}}\right) ^{\Delta }
\end{equation*}%
with
\begin{equation*}
\Delta =\Delta _{+}=a=\frac{1}{2}\left( d-1+\sqrt{(d-1)^{2}+4m^{2}\tilde{R}%
^{2}}\right)
\end{equation*}%
If we consider the conformally flat coordinates (\ref{metric: Euclid. conf.
flat}) in dS$^{d}$ the equations defining the Bulk to Boundary propagator
become

\begin{equation*}
\left( \square _{x}+m^{2}\right) K(x,\bar{y})=0
\end{equation*}%
We impose Dirichlet boundary conditions, $K \rightarrow \delta (x - \bar{y})$
as $x$ approaches the boundary of a spherical cap. The cap is then continued
to a hemisphere, and analytically continued to $\theta = {\frac{\pi}{2}} + i
t$. In our conformal coordinates for the Lorentzian section, $%
t\rightarrow\infty$, corresponds to $x^0 \rightarrow 0$. In this limit
\begin{equation*}
K(\bar{x},x_{0};\bar{y})\rightarrow C_+ (x_{0})^{((d-1)-\Delta_+ )}\delta
^{d}(\bar{x}-\bar{y}) + C_- (x_{0})^{((d-1)-\Delta_- )}\delta ^{d}(\bar{x}-%
\bar{y}),~\text{for }x_{0}\rightarrow 0
\end{equation*}

with%
\begin{equation*}
\Delta _{\pm}=a=\frac{1}{2}\left( d-1\pm \sqrt{(d-1)^{2}-4m^{2}R^{2}}\right)
\end{equation*}

\subsection{Boundary Two-point Function: dS/AdS}

The boundary two point function for AdS$^{d}$ in the Poincare patch as given
for example in \cite{Aharony:1999ti} is

\begin{equation*}
\langle 0|O(\bar{x})O(\bar{y})|0\rangle =\frac{\delta Z}{\delta \phi (\bar{x}%
)\delta \phi (\bar{y})}=C\frac{1}{(\bar{x}-\bar{y})^{2\Delta }}
\end{equation*}%
with%
\begin{equation*}
\Delta =\Delta _{+}=a=\frac{1}{2}\left( d-1+\sqrt{(d-1)^{2}+4m^{2}\tilde{R}%
^{2}}\right)
\end{equation*}

For dS$^{d}$ in the conformally flat coordinates (\ref{metric: Euclid. conf.
flat}) we have%
\begin{equation*}
\frac{\delta \Psi _{0}[h_{ij},\phi _{0}]}{\delta \phi _{0}(\bar{x})\delta
\phi _{0}(\bar{y})}=C_+ \frac{1}{(\bar{x}-\bar{y})^{2\Delta _{+}}} + C_-
\frac{1}{(\bar{x}-\bar{y})^{2\Delta _{-}}}
\end{equation*}%
with%
\begin{equation*}
\Delta _{\pm}=a=\frac{1}{2}\left( d-1 \pm \sqrt{(d-1)^{2}-4m^{2}R^{2}}\right)
\end{equation*}

\section{General Structure of the Computation\label{Sect.:General Structure
of the Computation}}

In this section we want to give a general description of the calculation we
will perform for three specific models.

\begin{figure}[tbp]
\begin{minipage}[t]{7cm}
\begin{center}
\includegraphics[width=7.0cm,clip]{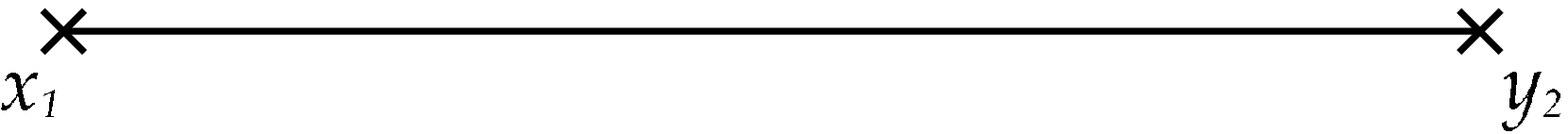}
\caption[Short caption for figure 1]{\textbf{Tree diagram.} {\em This
diagram represent the tree-level contribution to the Wave Function of the
Universe (WFU). The points $x_1$ and $x_2$ are on the boundary.}}
\label{treediag}
\end{center}
\end{minipage}
\hfill
\begin{minipage}[t]{7cm}
\begin{center}
\includegraphics[width=7.0cm,clip]{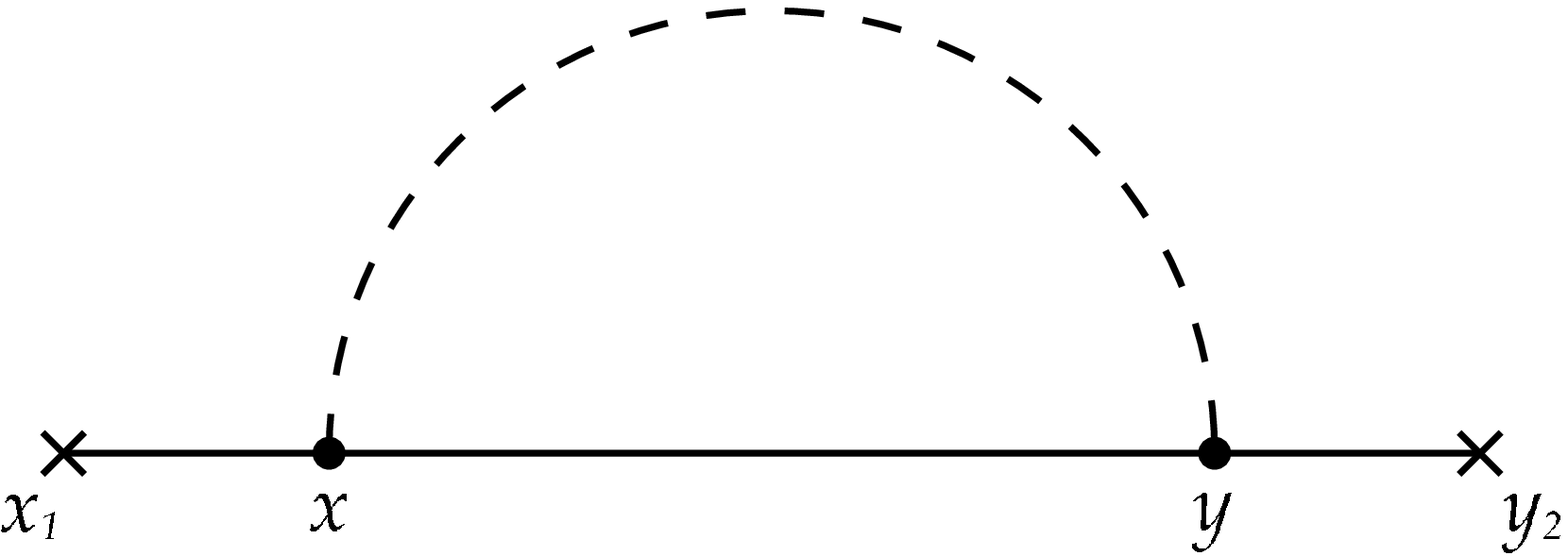}
\caption[Short caption for figure 2]{\textbf{1-loop diagram.} {\em 1-Loop
contribution to the WFU. The points $x_1$ and $x_2$ are on the boundary
while $x$ and $y$ are bulk points.}} \label{1loopdiag}
\end{center}
\end{minipage}
\end{figure}

As we have already discussed in Section \ref{Sect.: Wave Function of the
Universe} we are interesting in computing at \textit{1-loop} the Wave
Function of the Universe (WFU)%
\begin{equation*}
\Psi \lbrack h_{ij},\phi _{0}]=\int_{C}[dg][d\phi ]e^{-I[g,\phi ]}~
\end{equation*}%
for the models described in Section \ref{Sect.: Models}. The \textit{%
tree-level} and \textit{1-loop} diagrams are represented respectively in
Fig. \ref{treediag} and Fig. \ref{1loopdiag}.

Given the WFU we want to find the ``boundary two-point function"%
\begin{equation}
\frac{\delta \Psi \lbrack h_{ij},\phi _{0}]}{\delta \phi _{0}(\bar{x})\delta
\phi _{0}(\bar{y})}  \label{Second derivative
WFU}
\end{equation}%
this will provide us with a gauge invariant definition of the renormalized
mass.

We consider a general action of the form%
\begin{equation*}
S=\int d^{d}x\sqrt{g}\left( \phi _{A}\triangle \phi _{A}+\phi _{B}\triangle
\phi _{B}+\phi _{C}\triangle \phi _{C}\right) +\lambda \sqrt{g}\phi _{A}\phi
_{B}\phi _{C}
\end{equation*}%
where%
\begin{equation*}
S_{0}=\int d^{d}x\sqrt{g}\phi _{\alpha }\triangle \phi _{\alpha },~~\alpha
=A,B,C
\end{equation*}%
is the quadratic part of the action i.e.%
\begin{equation*}
S_{0}=\int d^{d}x\sqrt{g}\frac{1}{2}\left[ (\partial \phi
_{A})^{2}+m_{A}^{2}\phi _{A}^{2}\right]
\end{equation*}%
for a scalar field and%
\begin{equation*}
S_{0}=S_{M}+S_{\partial M}=\int_{M}d^{d}x\sqrt{g}\bar{\psi}\left(
D\!\!\!\!/-m\right) \psi +\int_{\partial M}d^{d}x~\sqrt{h}\bar{\psi}\psi
\end{equation*}%
for a spinor field.

In the WFU we are integrating over fields with the following boundary
conditions%
\begin{equation*}
\phi _{\alpha }|_{\Sigma }=\phi _{\alpha 0},~~\alpha =A,B,C
\end{equation*}%
where by the symbol $\phi _{\alpha }|_{\Sigma }$ we mean the field evaluated
on the boundary of the Euclidean spherical cap. To impose the boundary
condition we decompose the field in the following way%
\begin{equation*}
\phi _{\alpha }=\phi _{\alpha 1}+\phi _{\alpha 2}
\end{equation*}%
with%
\begin{equation*}
\phi _{\alpha 1}|_{\Sigma }=\phi _{\alpha 0},~~\phi _{\alpha 2}|_{\Sigma }=0
\end{equation*}%
The field $\phi _{\alpha 1}$ is the solution of the free wave equation with
Dirichlet boundary conditions, and can be written in terms of the
appropriate Bulk to Boundary propagator
\begin{equation*}
\phi _{\alpha 1}=K_{\alpha }\circ \phi _{\alpha 0}=\int_{\Sigma }d\bar{y}%
~K_{\alpha }(\bar{x},x_{0};\bar{y})\phi _{\alpha 0}(\bar{y}),~~\bar{y}\in
\Sigma ,~~\alpha =A,B,C
\end{equation*}%
described in the Sections \ref{Sect.: B-B prop. scalar} and \ref{Sect.: B-B
prop. spinor}.

To compute the \textit{1-loop }correction to the "boundary two-point
function" (\ref{Second derivative WFU}) we have to evaluate the terms in $%
\Psi \lbrack h_{ij},\phi _{0}]$ that are quadratic both in $\phi _{0}$ and
in the coupling constant $\lambda $. Expanding the path integral we have%
\begin{eqnarray*}
\Psi &=&\int [d\phi _{A}][d\phi _{B}]~[d\phi _{C}]e^{-S_{0}[\phi _{A},\phi
_{B},\phi _{C}]-\int d^{d}x~\sqrt{g(x)}\lambda \phi _{A}\phi _{B}\phi _{C}}
\\
&=&\int [d\phi _{A}][d\phi _{B}]~[d\phi _{C}]e^{-S_{0}[\phi _{A},\phi
_{B},\phi _{C}]}\left[ 1-\lambda \int d^{d}x~\sqrt{g(x)}\phi _{A}(x)\phi
_{B}(x)\phi _{C}(x)\right. \\
&&\left. +\frac{\lambda ^{2}}{2}\int d^{d}x\int d^{d}y~\sqrt{g(x)}\sqrt{g(y)}%
\phi _{A}(x)\phi _{B}(x)\phi _{C}(x)\phi _{A}(y)\phi _{B}(y)\phi
_{C}(y)+O(\lambda ^{3})\right]
\end{eqnarray*}%
where we indicated with $S_{0}[\phi _{A},\phi _{B},\phi _{C}]$ the quadratic
part of the action.

The terms quadratic in $\phi _{\alpha 0}~\alpha =A,B,C$ come from the
expansion of the term%
\begin{equation*}
\phi _{A}(x)\phi _{B}(x)\phi _{C}(x)\phi _{A}(y)\phi _{B}(y)\phi _{C}(y)
\end{equation*}%
we have%
\begin{eqnarray*}
&&\phi _{A}(x)\phi _{B}(x)\phi _{C}(x)\phi _{A}(y)\phi _{B}(y)\phi _{C}(y) \\
&=&\phi _{A1}(x)\phi _{A1}(y)\left[ \phi _{B2}(x)\phi _{B2}(y)\phi
_{C2}(x)\phi _{C2}(y)\right] \\
&&+\cdots
\end{eqnarray*}

We will compute only the correction to the two-point function of the field $%
\phi _{A}$. The part of the path integral relevant to this calculation is
\begin{eqnarray*}
\Psi _{\text{\textit{1-loop}}}^{\text{\textit{A}}}[\phi _{A0}] &=&\int
[d\phi _{B}]~[d\phi _{C}]e^{-\int d^{d}x\sqrt{g}\phi _{B}\triangle \phi
_{B}-\int d^{d}x\sqrt{g}\phi _{C}\triangle \phi _{C}} \\
&&\times \frac{\lambda ^{2}}{2}\int d^{d}x\int d^{d}y~\sqrt{g(x)}\sqrt{g(y)}%
\phi _{A1}(x)\phi _{A1}(y)\left[ \phi _{B2}(x)\phi _{B2}(y)\phi _{C2}(x)\phi
_{C2}(y)\right]
\end{eqnarray*}%
The only parts of the fields that fluctuate in the path integral are $\phi
_{\alpha 2}$, in fact $\phi _{\alpha 1}$ is fixed by the boundary
conditions. For this reason the measure of integration is given by%
\begin{equation*}
\lbrack d\phi _{B}]~[d\phi _{C}]=[d\phi _{B2}]~[d\phi _{C2}]
\end{equation*}

Standard manipulation give us the following expression for the path integral%
\begin{eqnarray*}
\Psi _{\text{\textit{1-loop}}}^{\text{\textit{A}}}[\phi _{A0}] &=&\int
[d\phi _{B2}]~[d\phi _{C2}]e^{-\int d^{d}x\sqrt{g}\phi _{B}\triangle \phi
_{B}-\int d^{d}x\sqrt{g}\phi _{C}\triangle \phi _{C}} \\
&&\times \frac{\lambda ^{2}}{2}\int d^{d}x\int d^{d}y~\sqrt{g(x)}\sqrt{g(y)}%
\phi _{A1}(x)\phi _{A1}(y)[\phi _{B2}(x)\phi _{B2}(y)\phi _{C2}(x)\phi
_{C2}(y)] \\
&=&\frac{\lambda ^{2}}{2}\int d^{d}x\int d^{d}y~\sqrt{g(x)}\sqrt{g(y)}\phi
_{A1}(x)\phi _{A1}(y)\langle E|\phi _{B2}(x)\phi _{B2}(y)\phi _{C2}(x)\phi
_{C2}(y)|E\rangle \\
&=&\frac{\lambda ^{2}}{2}\int d^{d}x\int d^{d}y~\sqrt{g(x)}\sqrt{g(y)}\phi
_{A1}(x)\phi _{A1}(y)\langle E|\phi _{B}(x)\phi _{B}(y)\phi _{C}(x)\phi
_{C}(y)|E\rangle \\
&=&\frac{\lambda ^{2}}{2}\int d^{d}x\int d^{d}y~\sqrt{g(x)}\sqrt{g(y)}\phi
_{A1}(x)\phi _{A1}(y)G_{B}(x,y)G_{C}(x,y) \\
&=&\frac{\lambda ^{2}}{2}\int d^{d}x\int d^{d}y~\sqrt{g(x)}\sqrt{g(y)}%
\int_{\Sigma }d\bar{x}_{1}~K_{A}(x;\bar{x}_{1})\phi _{A0}(\bar{x}_{1}) \\
&&\times \int_{\Sigma }d\bar{x}_{2}~K_{A}(y;\bar{x}_{2})\phi _{A0}(\bar{x}%
_{2})G_{B}(x,y)G_{C}(x,y)
\end{eqnarray*}

The boundary two-point function at \textit{1-loop} is given by%
\begin{eqnarray*}
\frac{\delta \Psi ^{\text{\textit{A}}}}{\delta \phi _{A0}(\bar{x}_{1})\delta
\phi _{A0}(\bar{x}_{2})} &=&C\frac{1}{(\bar{x}_{1}-\bar{x}_{2})^{2\Delta }}
\\
&&+\frac{\lambda ^{2}}{2}\int d^{d}x\int d^{d}y~\sqrt{g(x)}\sqrt{g(y)}%
~K_{A}(x;\bar{x}_{1})G_{B}(x,y)G_{C}(x,y)K_{A}(y;\bar{x}_{2})
\end{eqnarray*}

We have similar expressions for the boundary two-point functions of the
others fields $\phi _{B},~\phi _{C}$.

\section{Models\label{Sect.: Models}}

We have computed the \textit{1-loop} boundary two point function for the
following models:

\textbf{Scalar Fields with Cubic Interaction}%
\begin{equation*}
S=\int d^{d}x\sqrt{g}\frac{1}{2}\left[ (\partial \phi )^{2}+m^{2}\phi
^{2}+(\partial \phi _{1})^{2}+m^{2}\phi _{1}^{2}+(\partial \phi _{0})^{2}%
\right] +\sqrt{g}\lambda \phi \phi _{0}\phi _{1}
\end{equation*}%
where the field $\phi _{0}$ is massless.

\textbf{Scalar Fields with Derivative Couplings}

\begin{equation*}
S=\int d^{d}x\sqrt{g}\frac{1}{2}\left[ (\partial \phi _{A})^{2}+m^{2}\phi
_{A}^{2}+(\partial \phi _{B})^{2}+m^{2}\phi _{B}^{2}+(\partial \phi )^{2}%
\right] +\sqrt{g}\lambda \phi g^{\mu \nu }\partial _{\mu }\phi _{A}\partial
_{\nu }\phi _{B}
\end{equation*}%
where the field $\phi $ is massless.

\textbf{Spinor Field with Derivative Coupling}

\begin{equation*}
S_{0}+S_{I}=\int d^{d}x\frac{1}{2}\sqrt{g}(\partial \phi )^{2}+\int_{M}d^{d}x%
\sqrt{g}\bar{\psi}(D\!\!\!\!/-m)\psi +\int_{\partial M}d^{d}x~\sqrt{h}\bar{%
\psi}\psi +\int_{M}d^{d}x~\sqrt{g}\lambda \partial _{a}\phi \bar{\psi}\Gamma
^{a}\psi
\end{equation*}%
where the field $\phi $ is massless. The surface term for the fermions is
explained in \cite{Arutyunov:1998ve},\cite{Rashkov:1999ji},\cite%
{Henningson:1998cd}.

We have chosen these models in order to see whether the fact that the
massless boson is derivatively coupled effects the IR divergence, and to
study the effect of fermion chirality. In the conclusions we will discuss
the issues that these results raise for the analogous calculations in
quantum supergravity.

\section{1-loop Computation: Scalar Fields with Cubic Interaction}

In this section we will compute the \textit{1-loop} boundary two point
function for the massive field $\phi $ interacting with a massive scalar
field $\phi _{1}$ and a massless scalar field $\phi _{0}$. The lagrangian is%
\begin{equation*}
\mathcal{L}=\sqrt{g}\frac{1}{2}\left[ (\partial \phi )^{2}+m^{2}\phi
^{2}+(\partial \phi _{1})^{2}+m^{2}\phi _{1}^{2}+(\partial \phi _{0})^{2} %
\right] +\sqrt{g}\lambda \phi \phi _{0}\phi _{1}
\end{equation*}

The asymptotic expansions of both the bulk and bulk to boundary propagators,
at large Lorentzian time and space-like separation, contain terms with both
powers $(x_0)^{\Delta_{\pm}}$. For the principal series, these powers differ
in the sign of their imaginary part. We have found that the most divergent
terms as $x_0 \rightarrow 0$ come from products of terms from individual
propagators that all have the same power of $x_0$. We call these the \textit{%
pure} terms. Mixed terms have rapidly oscillating phases, which lead to more
convergent integrals. We will find that in this model the pure terms look
like the tree level results, but with a divergent correction to the mass.
The mixed terms are sub-leading, and do not have the same form as the tree
level result. We will explicitly show only our results for the pure terms.

As explained in Section \ref{Sect.:General Structure of the Computation} the
\textit{1-loop} correction to the boundary two-point function%
\begin{equation*}
\frac{\delta \Psi _{\text{\textit{1-loop}}}}{\delta \phi _{0}(\bar{x}%
_{1})\delta \phi _{0}(\bar{x}_{2})}=G_{\text{\textit{1-loop}}}(\bar{x}_{1},%
\bar{x}_{2})
\end{equation*}

is given by%
\begin{eqnarray*}
G_{\text{\textit{1-loop}}}(\bar{x}_{1},\bar{x}_{2}) &=&\frac{\lambda ^{2}}{2}%
\int d^{d}x\int d^{d}y~\sqrt{g(x)}\sqrt{g(y)}~K(x;\bar{x}%
_{1})G_{1}(x,y)G_{0}(x,y)K(y;\bar{x}_{2}) \\
&=&\frac{\lambda ^{2}}{2}\int d^{d-1}\bar{x}\int d^{d-1}\bar{y}\int
dx_{0}\int dy_{0}~\frac{1}{x_{0}^{d}}\frac{1}{y_{0}^{d}}~K(x;\bar{x}%
_{1})G_{1}(x,y)G_{0}(x,y)K(y;\bar{x}_{2})
\end{eqnarray*}%
In principle, the bulk propagators in these equations should satisfy
(vanishing) Dirichlet boundary conditions at a fixed global time, $T$. We
have seen that in conformal coordinates this corresponds to an $x^{0}$
dependent Dirichlet boundary condition on a sphere in $\mathbf{x}$ space, as
well as an upper cut-off $x_{max}^{0}\sim -R^{2}/T$. The IR divergences will
come from the regions of maximal spatial geodesic size, and, because of the
Dirichlet boundary conditions, from regions where the two integrated bulk
points are far from the spatial boundary sphere. Thus considering only the
leading IR divergent part of the answer, we can use the usual Euclidean
vacuum Green's function (without Dirichlet boundary conditions) and
approximate it by its asymptotic form at large geodesic distance:%
\begin{eqnarray*}
G_{\text{\textit{1-loop}}}^{\text{\textit{IR}}}(\bar{x}_{1},\bar{x}_{2})
&\sim &\frac{\lambda ^{2}}{2}\int d^{d-1}\bar{x}\int d^{d-1}\bar{y}\int
dx_{0}\int dy_{0}~\frac{1}{x_{0}^{d}}\frac{1}{y_{0}^{d}}%
~(x_{0}y_{0})^{((d-1)-\Delta _{\pm })} \\
&&\times \delta ^{d-1}(\bar{x}-\bar{x}_{1})G_{1}(x,y)G_{0}(x,y)\delta ^{d-1}(%
\bar{y}-\bar{x}_{2}) \\
&=&\frac{\lambda ^{2}}{2}\int dx_{0}\int dy_{0}~\frac{1}{x_{0}^{d}}\frac{1}{%
y_{0}^{d}}~(x_{0}y_{0})^{((d-1)-\Delta _{\pm })}G_{0}(\bar{x}_{1},x_{0};\bar{%
x}_{2},y_{0})G_{1}(\bar{x}_{1},x_{0};\bar{x}_{2},y_{0}) \\
&\sim &\frac{\lambda ^{2}}{2}\int_{\alpha }^{\epsilon }dx_{0}\int_{\beta
}^{\epsilon }dy_{0}~\frac{1}{x_{0}}\frac{1}{y_{0}}~C_{0}C_{-}\ln \left(
\frac{(\bar{x}-\bar{y})^{2}}{x_{0}y_{0}}\right) \left( \frac{1}{(\bar{x}-%
\bar{y})^{2}}\right) ^{\Delta _{\pm }}
\end{eqnarray*}%
Here we used the fact that bulk to boundary propagators satisfy

\begin{equation*}
K(\bar{x},x_{0};\bar{y})\rightarrow C_{+}(x_{0})^{((d-1)-\Delta _{+})}\delta
^{d}(\bar{x}-\bar{y})+C_{-}(x_{0})^{((d-1)-\Delta _{-})}\delta ^{d}(\bar{x}-%
\bar{y}),~\text{for }x_{0}\rightarrow 0
\end{equation*}%
explained in Section \ref{Sect.: B-B prop. scalar} and the asymptotic
expansion (\ref{2-point fnct.: scalar, asympt. exp.}), (\ref{2-point fnct.:
m=0, scalar, asympt. exp.}) for the bulk two-point functions\footnote{%
In tree level calculations involving two bulk to boundary propagators, only
one of them can be replaced by a $\delta $ function, since the other ends up
evaluated at separated points. The powers of $x^{0}$ that would set it equal
to zero are part of the renormalization factor that defines the limiting
boundary two point function. In our calculation, both bulk to boundary
propagators are legitimately replaced by $\delta $ functions.} .

Integrating in $x_{0}$ and $y_{0}$ and keeping the leading part in $\epsilon
\rightarrow 0$ we find%
\begin{eqnarray}
G_{\text{\textit{1-loop}}}^{\text{\textit{IR}}}(\bar{x}_{1},\bar{x}_{2})
&\sim &\frac{\lambda ^{2}}{2}\frac{1}{(\bar{x}_{1}-\bar{x}_{2})^{2\Delta
_{\pm }}}\times \left( \ln \left( \frac{(\bar{x}_{1}-\bar{x}_{2})^{2}}{%
\epsilon }\right) \right) ^{3}  \notag \\
&&+\text{Subleading terms in }\epsilon   \notag
\end{eqnarray}

\section{1-loop Computation: Scalar Fields with Derivative Coupling}

In this section we will compute the \textit{1-loop} boundary two points
function for the massive scalar field $\phi $ derivatively coupled to a
massless scalar field $\phi _{A}$ and a massive scalar field $\phi _{B}$.
The action is

\begin{equation*}
S=\int d^{d}x\sqrt{g}\frac{1}{2}\left[ (\partial \phi )^{2}+m^{2}\phi
^{2}+(\partial \phi _{B})^{2}+m^{2}\phi _{B}^{2}+(\partial \phi _{A})^{2}%
\right] +\sqrt{g}\lambda \phi g^{\mu \nu }\partial _{\mu }\phi _{A}\partial
_{\nu }\phi _{B}
\end{equation*}

Following the general lines of the computation done in Section \ref%
{Sect.:General Structure of the Computation} we find for the \textit{1-loop}
WFU

\begin{eqnarray*}
\Psi _{\text{\textit{1-loop}}} &=&\int [d\phi _{B2}]~[d\phi _{C2}]e^{-\int
d^{d}x\sqrt{g}\frac{1}{2}\left[ (\partial \phi )^{2}+m^{2}\phi
^{2}+(\partial \phi _{B})^{2}+m^{2}\phi _{B}^{2}+(\partial \phi _{A})^{2}%
\right] } \\
&&\times \frac{\lambda ^{2}}{2}\int d^{d}x\int d^{d}y~\sqrt{g(x)}\sqrt{g(y)}
\\
&&\times (\phi (x)g^{\mu \nu }(x)\partial _{\mu }\phi _{A}(x)\partial _{\nu
}\phi _{B}(x))(\phi (y)g^{\rho \lambda }(y)\partial _{\rho }\phi
_{A}(y)\partial _{\lambda }\phi _{B}(y)) \\
&=&\frac{\lambda ^{2}}{2}\int d^{d}x\int d^{d}y~\sqrt{g(x)}\sqrt{g(y)}\phi
_{1}(x)\phi _{1}(y) \\
&&\times g^{\mu \nu }(x)g^{\rho \lambda }(y)\partial _{\mu }^{x}\partial
_{\rho }^{y}G_{A}(x,y)\partial _{\nu }^{x}\partial _{\lambda }^{y}G_{B}(x,y)
\\
&=&\frac{\lambda ^{2}}{2}\int d^{d}x\int d^{d}y~\sqrt{g(x)}\sqrt{g(y)}%
\int_{\Sigma }d\bar{x}_{1}~K_{A}(x;\bar{x}_{1})\phi _{0}(\bar{x}%
_{1})\int_{\Sigma }d\bar{x}_{2}~K_{A}(y;\bar{x}_{2})\phi _{0}(\bar{x}_{2}) \\
&&\times g^{\mu \nu }(x)g^{\rho \lambda }(y)\partial _{\mu }^{x}\partial
_{\rho }^{y}G_{A}(x,y)\partial _{\nu }^{x}\partial _{\lambda }^{y}G_{B}(x,y)
\end{eqnarray*}

The \textit{1-loop} two point function is%
\begin{equation*}
\frac{\delta \Psi _{\text{\textit{1-loop}}}}{\delta \phi _{0}(\bar{x}%
_{1})\delta \phi _{0}(\bar{x}_{2})}=G_{\text{\textit{1-loop}}}(\bar{x}_{1},%
\bar{x}_{2})
\end{equation*}

Considering only the leading IR divergent part we have%
\begin{eqnarray*}
G_{\text{\textit{1-loop}}}^{\text{\textit{IR}}}(\bar{x}_{1},\bar{x}_{2})
&\sim &\frac{\lambda ^{2}}{2}\int dx_{0}\int dy_{0}~\frac{1}{x_{0}^{d}}\frac{%
1}{y_{0}^{d}}~(x_{0}y_{0})^{((d-1)-\Delta _{\pm })} \\
&&\times g^{\mu \nu }(x)g^{\rho \lambda }(y)\partial _{\mu }^{x}\partial
_{\rho }^{y}C_{0}C_{-}\ln \left( \frac{(\bar{x}_{1}-\bar{x}_{2})^{2}}{%
x_{0}y_{0}}\right) \partial _{\nu }^{x}\partial _{\lambda }^{y}\left( \frac{%
x_{0}y_{0}}{(\bar{x}_{1}-\bar{x}_{2})^{2}}\right) ^{\Delta _{\pm }} \\
&=&\frac{\lambda ^{2}}{2}\int dx_{0}\int dy_{0}~\frac{1}{x_{0}^{d}}\frac{1}{%
y_{0}^{d}}~x_{0}^{2}y_{0}^{2}(x_{0}y_{0})^{((d-1)-\Delta _{\pm })}\partial
_{\mu }^{x}\partial _{\rho }^{y}C_{0}C_{-} \\
&&\times \ln \left( \frac{(\bar{x}_{1}-\bar{x}_{2})^{2}}{x_{0}y_{0}}\right)
\partial _{\mu }^{x}\partial _{\rho }^{y}\left( \frac{x_{0}y_{0}}{(\bar{x}%
_{1}-\bar{x}_{2})^{2}}\right) ^{\Delta _{\pm }} \\
&=&\frac{\lambda ^{2}}{2}\int dx_{0}\int dy_{0}~\frac{1}{x_{0}^{d}}\frac{1}{%
y_{0}^{d}}~x_{0}^{2}y_{0}^{2}(x_{0}y_{0})^{((d-1)-\Delta _{\pm })}\partial
_{i}^{x}\partial _{j}^{y}C_{0}C_{-} \\
&&\times \ln \left( (\bar{x}_{1}-\bar{x}_{2})^{2}\right) \partial
_{i}^{x}\partial _{j}^{y}\left( \frac{x_{0}y_{0}}{(\bar{x}_{1}-\bar{x}%
_{2})^{2}}\right) ^{\Delta _{\pm }} \\
&=&\frac{\lambda ^{2}}{2}\int dx_{0}\int dy_{0}~x_{0}^{1}y_{0}^{1}(-4\Delta
(3+2\Delta ))C_{0}C_{-}\left( \frac{1}{(\bar{x}_{1}-\bar{x}_{2})^{2}}\right)
^{2+\Delta _{\pm }}
\end{eqnarray*}%
where we used the bulk to boundary propagators property explained in Section %
\ref{Sect.: B-B prop. scalar} and the asymptotic expansion (\ref{2-point
fnct.: scalar, asympt. exp.}), (\ref{2-point fnct.: m=0, scalar, asympt.
exp.}) for the bulk two-point functions. Furthermore we used the fact that%
\begin{equation*}
\partial _{0}^{x}\partial _{0}^{y}\left( \ln \frac{(\bar{x}-\bar{y})^{2}}{%
x_{0}y_{0}}\right) =0,~~\partial _{0}^{x}\partial _{j}^{y}\left( \ln \frac{(%
\bar{x}-\bar{y})^{2}}{x_{0}y_{0}}\right) =0
\end{equation*}%
with $i,j=1,\ldots ,d$ .

Doing the integrals and keeping the leading parts in $\epsilon \rightarrow 0$
we find%
\begin{eqnarray}
G_{\text{\textit{1-loop}}}^{\text{\textit{IR}}}(\bar{x}_{1},\bar{x}_{2})
&\sim &(\epsilon )^{4}\left( \frac{1}{(\bar{x}_{1}-\bar{x}_{2})^{2}}\right)
^{2+\Delta _{-}}
\label{Bdry 2-point fnct 1-loop: dS, 3 scalar deriv. coupl.} \\
&&+\text{Subleading terms in }\epsilon   \notag
\end{eqnarray}

\section{1-loop Computation: Spinor Field with Derivative Coupling}

In this last section we will evaluate the \textit{1-loop} boundary two-point
function for a spinor field $\psi $ derivatively coupled to a massless
scalar field $\phi $. The action in the tangent frame is

\begin{equation*}
S_{0}+S_{I}=\int_{M}d^{d}x\frac{1}{2}\sqrt{g}(\partial \phi
)^{2}+\int_{M}d^{d}x\sqrt{g}\bar{\psi}(D\!\!\!\!/-m)\psi +\int_{\partial
M}d^{d}x~\sqrt{h}\bar{\psi}\psi +\int_{M}d^{d}x~\lambda \sqrt{g}\partial
_{a}\phi \bar{\psi}\Gamma ^{a}\psi
\end{equation*}%
The surface term for the fermions is explained in \cite{Arutyunov:1998ve},%
\cite{Rashkov:1999ji},\cite{Henningson:1998cd}.

More specifically we are using the metric%
\begin{equation*}
ds^{2}=-\frac{1}{x_{0}^{2}}(dx^{0}dx^{0}+d\bar{x}\cdot d\bar{x})=-\frac{1}{%
x_{0}^{2}}(dx^{0}dx^{0}+dx_{i}dx_{i})
\end{equation*}

and the vielbein $e_{\mu }^{a},~a=0,\ldots ,d-1$ such that $g_{\mu \nu }=$ $%
e_{\mu }^{a}e_{\nu }^{b}\eta _{ab}$. The explicit form of the vielbein and
is inverse is%
\begin{eqnarray*}
e_{\mu }^{a} &=&\frac{1}{x_{0}}\delta _{\mu }^{a} \\
e_{a}^{\mu } &=&x_{0}\delta _{a}^{\mu }
\end{eqnarray*}%
the spin connection has the form%
\begin{equation*}
\omega _{i}^{0j}=\omega _{i}^{j0}=\frac{1}{x_{0}}\delta _{i}^{j}
\end{equation*}%
and all other component vanishing. The Dirac operator is given by%
\begin{equation*}
D\!\!\!\!/=e_{a}^{\mu }(\partial _{\mu }+\frac{1}{2}\omega _{\mu
}^{bc}\Sigma _{bc})=x_{0}\Gamma ^{0}\partial _{0}+x_{0}\bar{\Gamma}\cdot
\nabla -\frac{d-1}{2}\Gamma ^{0}
\end{equation*}%
where $\Gamma ^{a}=(\Gamma ^{0},\Gamma ^{i})=(\Gamma ^{0},\bar{\Gamma})$
satisfy $\{\Gamma ^{a},\Gamma ^{b}\}=2\eta ^{ab}$ and $\partial _{\mu
}=(\partial _{0},\partial _{i})=(\partial _{0},\nabla )$.

The explicit form of the interacting term is
\begin{equation*}
\mathcal{L}_{I}=\lambda \sqrt{g}\partial _{a}\phi \bar{\psi}\Gamma ^{a}\psi
=\lambda \sqrt{g}e_{a}^{\mu }\partial _{\mu }\phi \bar{\psi}\Gamma ^{a}\psi
=\lambda \sqrt{g}x_{0}\delta _{a}^{\mu }\partial _{\mu }\phi \bar{\psi}%
\Gamma ^{a}\psi
\end{equation*}

Again following the same reasoning of Section \ref{Sect.:General Structure
of the Computation} we find for the \textit{1-loop} WFU%
\begin{eqnarray*}
\Psi _{\text{\textit{1-loop}}} &=&\int [d\psi ]~[d\bar{\psi}]e^{-\left(
\int_{M}d^{d}x\frac{1}{2}\sqrt{g}(\partial \phi )^{2}+\int_{M}d^{d}x\sqrt{g}%
\bar{\psi}(D\!\!\!\!/-m)\psi +\int_{\partial M}d^{d}x~\sqrt{h}\bar{\psi}\psi
+\int_{M}d^{d}x~\lambda \sqrt{g}\partial _{a}\phi \bar{\psi}\Gamma ^{a}\psi
\right) } \\
&=&\int [d\psi ]~[d\bar{\psi}]e^{-S_{0}}\int d^{d}x\int d^{d}y~\sqrt{g(x)}%
\sqrt{g(y)} \\
&&\times \frac{\lambda ^{2}}{2}\left( \partial _{a}\phi (x)\bar{\psi}%
(x)\Gamma ^{a}\psi (x)\right) \left( \partial _{b}\phi (y)\bar{\psi}%
(y)\Gamma ^{b}\psi (y)\right)  \\
&=&\frac{\lambda ^{2}}{2}\int d^{d}x\int d^{d}y~\sqrt{g(x)}\sqrt{g(y)}\bar{%
\psi}_{1}(x)\langle E|\partial _{a}\phi (x)\Gamma ^{a}\psi (x)\partial
_{b}\phi (y)\bar{\psi}(y)\Gamma ^{b}|E\rangle \psi _{1}(y) \\
&=&\frac{\lambda ^{2}}{2}\int d^{d}x\int d^{d}y~\sqrt{g(x)}\sqrt{g(y)}\bar{%
\psi}_{1}(x)\Gamma ^{a}S(x,y)\Gamma ^{b}\partial _{a}^{x}\partial
_{b}^{y}G_{0}(x,y)\psi _{1}(y) \\
&=&\frac{\lambda ^{2}}{2}\int d^{d}x\int d^{d}y~\sqrt{g(x)}\sqrt{g(y)}\int
d^{d-1}\bar{x}~\bar{\psi}_{0}(\bar{x})K(y,\bar{x})\Gamma ^{a}S(x,y)\Gamma
^{b}\partial _{a}^{x}\partial _{b}^{y}G_{0}(x,y) \\
&&\times \int d^{d-1}\bar{x}~K(x,\bar{x})\psi _{0}(\bar{x})
\end{eqnarray*}

taking the limit $x_{0}\rightarrow 0,~y_{0}\rightarrow 0$ we find the
leading IR part of $\Psi _{\text{\textit{1-loop}}}$
\begin{eqnarray*}
\Psi _{\text{\textit{1-loop}}}^{\text{\textit{IR}}} &\sim &\frac{\lambda ^{2}%
}{2}\int d^{d}x\int d^{d}y~\frac{1}{x_{0}^{d}}\frac{1}{y_{0}^{d}}\bar{\psi}%
_{0+}(\bar{x}_{1})\Gamma ^{a}S(x,y)\Gamma ^{b}\partial _{a}^{x}\partial
_{b}^{y}G_{0}(x,y)\psi _{0-}(\bar{x}_{2}) \\
&\sim &\frac{\lambda ^{2}}{2}\int d^{d}x\int d^{d}y~\frac{1}{x_{0}^{d}}\frac{%
1}{y_{0}^{d}} \\
&&\times \bar{\psi}_{0+}(\bar{x}_{1})\Gamma ^{a}C_{-}C_{0}\left( \frac{%
x_{0}y_{0}}{(\bar{x}-\bar{y})^{2}}\right) ^{\Delta _{-}}\frac{\bar{\Gamma}%
\cdot (\bar{x}-\bar{y})}{\left\vert \bar{x}-\bar{y}\right\vert }\Gamma
^{b}\partial _{a}^{x}\partial _{b}^{y}\ln \left( \frac{(\bar{x}-\bar{y})^{2}%
}{x_{0}y_{0}}\right) \psi _{0-}(\bar{x}_{2}) \\
&=&\frac{\lambda ^{2}}{2}\int d^{d}x\int d^{d}y~\frac{1}{x_{0}^{d}}\frac{1}{%
y_{0}^{d}}(x_{0}y_{0})^{\Delta _{1}+1} \\
&&\times \bar{\psi}_{0+}(\bar{x}_{1})\Gamma ^{a}C_{-}C_{0}\left( \frac{%
x_{0}y_{0}}{(\bar{x}-\bar{y})^{2}}\right) ^{\Delta _{-}}\frac{\bar{\Gamma}%
\cdot (\bar{x}-\bar{y})}{\left\vert \bar{x}-\bar{y}\right\vert }\Gamma
^{b}\delta _{a}^{\mu }\partial _{\mu }^{x}\delta _{b}^{\nu }\partial _{\nu
}^{y}\ln \left( \frac{(\bar{x}-\bar{y})^{2}}{x_{0}y_{0}}\right) \psi _{0-}(%
\bar{x}_{2}) \\
&=&\frac{\lambda ^{2}}{2}\int d^{d}x\int d^{d}y~\frac{1}{x_{0}^{d}}\frac{1}{%
y_{0}^{d}}(x_{0}y_{0})^{\Delta _{-}+1} \\
&&\times \bar{\psi}_{0+}(\bar{x}_{1})\Gamma ^{a}C_{-}C_{0}\left( \frac{%
x_{0}y_{0}}{(\bar{x}-\bar{y})^{2}}\right) ^{\Delta _{-}}\frac{\bar{\Gamma}%
\cdot (\bar{x}-\bar{y})}{\left\vert \bar{x}-\bar{y}\right\vert }\Gamma
^{b}\delta _{a}^{i}\partial _{i}^{x}\delta _{b}^{j}\partial _{j}^{y}\ln
\left( (\bar{x}-\bar{y})^{2}\right) \psi _{0-}(\bar{x}_{2}) \\
&=&\frac{\lambda ^{2}}{2}C_{-}C_{0}\int d^{d}x\int d^{d}y~\frac{1}{x_{0}^{d}}%
\frac{1}{y_{0}^{d}}(x_{0}y_{0})^{\Delta _{-}+1} \\
&&\times \bar{\psi}_{0+}(\bar{x}_{1})\Gamma ^{i}\left( \frac{1}{(\bar{x}-%
\bar{y})^{2}}\right) ^{\Delta _{-}}\frac{\Gamma ^{k}(\bar{x}-\bar{y})_{k}}{%
\left\vert \bar{x}-\bar{y}\right\vert }\Gamma ^{j}\partial _{i}^{x}\partial
_{j}^{y}\ln \left( (\bar{x}-\bar{y})^{2}\right) \psi _{0-}(\bar{x}_{2})
\end{eqnarray*}%
where we used the bulk to boundary propagators property%
\begin{equation*}
\lim_{x_{0}\rightarrow 0}(x_{0})^{-\frac{d}{2}+m}\psi (x)=-c\psi _{0-}(\bar{x%
})
\end{equation*}%
\begin{equation*}
\lim_{x_{0}\rightarrow 0}(x_{0})^{-\frac{d}{2}+m}\bar{\psi}(x)=c\bar{\psi}%
_{0+}(\bar{x})
\end{equation*}%
explained in Appendix \ref{Sect.: B-B prop. spinor} and the asymptotic
expansion (\ref{2-point fnct.: spinor, asympt. exp.}), (\ref{2-point fnct.:
m=0, scalar, asympt. exp.}) for the bulk two-point functions. As in the
previous section we noticed that%
\begin{equation*}
\partial _{0}^{x}\partial _{0}^{y}\ln \left( \frac{(\bar{x}-\bar{y})^{2}}{%
x_{0}y_{0}}\right) =0,~~\partial _{0}^{x}\partial _{j}^{y}\ln \left( \frac{(%
\bar{x}-\bar{y})^{2}}{x_{0}y_{0}}\right) =0,~~i,j=1,\ldots ,d
\end{equation*}

The boundary two-point function at 1-loop is%
\begin{eqnarray*}
G_{\text{\textit{1-loop}}}(\bar{x}_{1},\bar{x}_{2}) &=&\frac{\delta \Psi _{%
\text{\textit{1-loop}}}^{\text{\textit{IR}}}}{\delta \bar{\psi}_{0+}(\bar{x}%
_{1})\delta \psi _{0-}(\bar{x}_{2})} \\
&=&\frac{\lambda ^{2}}{2}C_{0}C_{-}\int_{\alpha }^{\epsilon
}d^{0}x\int_{\beta }^{\epsilon }d^{0}y~(x_{0}y_{0})^{(\Delta _{-}+1-d)}~ \\
&&\times \Gamma ^{i}\left( \frac{1}{(\bar{x}_{1}-\bar{x}_{2})^{2}}\right)
^{\Delta _{-}}\frac{\Gamma ^{k}(\bar{x}_{1}-\bar{x}_{2})_{k}}{\left\vert
\bar{x}_{1}-\bar{x}_{2}\right\vert }\Gamma ^{j}\partial _{i}^{x}\partial
_{j}^{y}\ln \left( (\bar{x}_{1}-\bar{x}_{2})^{2}\right)
\end{eqnarray*}%
doing the integrals and keeping the leading terms in $\epsilon \rightarrow 0$
we find%
\begin{eqnarray}
G_{\text{\textit{1-loop}}}(\bar{x}_{1},\bar{x}_{2}) &\sim &\lambda
^{2}\epsilon ^{2(\Delta _{-}-d+2)}
\label{Bdry 2-point fnct 1-loop: dS, 1 scalar + 1 fermion} \\
&&\times \Gamma ^{i}\Gamma ^{k}\Gamma ^{j}\left( \frac{1}{(\bar{x}_{1}-\bar{x%
}_{2})^{2}}\right) ^{\Delta _{-}}\frac{(\bar{x}_{1}-\bar{x}_{2})_{k}}{%
\left\vert \bar{x}_{1}-\bar{x}_{2}\right\vert }\partial _{i}^{x}\partial
_{j}^{y}\ln \left( (\bar{x}_{1}-\bar{x}_{2})^{2}\right)   \notag \\
&&+\text{Subleading terms in }\epsilon   \notag
\end{eqnarray}

\section{Analysis Divergences}

\subsection{Three Massive Scalar Fields with Cubic Interaction in dS$^{d}$}

\subsubsection{Leading Terms}

We didn't perform explicitly the computation in this case but it is easy to
see that the leading IR divergent term (which is not in fact divergent in
this case) in the boundary two-point function has the following form up to a
constant

\begin{eqnarray*}
G_{\text{\textit{1-loop}}}^{\text{\textit{IR}}}(\bar{x}_{1},\bar{x}_{2})
&\sim &\frac{\lambda ^{2}}{2}\epsilon ^{2\Delta _{2}}\frac{1}{(\bar{x}_{1}-%
\bar{x}_{2})^{2\Delta _{2}}}\frac{1}{(\bar{x}_{1}-\bar{x}_{2})^{2\Delta _{1}}%
} \\
&&+\text{Subleading terms in }\epsilon
\end{eqnarray*}%
where $\Delta _{2},~\Delta _{1}$ correspond respectively to the fields $\phi
_{1}$ and $\phi _{2}$. In this expression we have kept only pure terms.
Other terms are no more divergent than these.

The leading IR term in $G_{\text{\textit{1-loop}}}^{\text{\textit{IR}}}(\bar{%
x}_{1},\bar{x}_{2})$ is proportional to%
\begin{equation*}
\epsilon ^{2\Delta _{2}}
\end{equation*}

We have%
\begin{equation*}
\Delta _{\pm }^{i}=\frac{1}{2}\left( d-1\pm \sqrt{%
(d-1)^{2}-4m_{i}^{2}R_{dS}^{2}}\right) =\frac{1}{2}(d-1)\left( 1\pm \sqrt{%
(1-\alpha _{i})}\right)
\end{equation*}%
with%
\begin{equation*}
\alpha _{i}=\left( \frac{2m_{i}R_{dS}}{d-1}\right) ^{2}
\end{equation*}%
So we immediately see that $G_{\text{\textit{1-loop}}}^{\text{\textit{IR}}}(%
\bar{x}_{1},\bar{x}_{2})$ is IR convergent for every $\alpha _{i}$ i.e. both
for the complementary and principal series, see Section \ref{Representions
of dS Group}.

This computation shows that in the case of massive fields there is no IR
divergence in the boundary two point function. This is in accord with naive
expectations.

\subsection{Two Massive and One Massless field in dS$^{d}$}

The leading IR term in $G_{\text{\textit{1-loop}}}(\bar{x}_{1},\bar{x}_{2})$
is proportional to%
\begin{equation*}
(\log \epsilon )^{3}
\end{equation*}

So in this case $G_{\text{\textit{1-loop}}}(\bar{x}_{1},\bar{x}_{2})$ is IR
divergent.

The analysis of divergences in the remaining case (\ref{Bdry 2-point fnct
1-loop: dS, 3 scalar deriv. coupl.}), (\ref{Bdry 2-point fnct 1-loop: dS, 1
scalar + 1 fermion}) is very similar and we will not repeat it. We want only
to remark that these cases are not IR\ divergent, due to the presence of
derivative couplings, as can be seen inspecting the power dependence of the $%
\epsilon $ cutoff.

\section{The Meaning of the Divergences}

To understand the meaning of the divergences we have found, we compare our
expressions to those obtained by perturbing the free massive theory by a
term ${\frac{1}{2}} \delta m^2 \phi^2$. That computation gives

\begin{equation*}
\delta m^{2}\int dx_{0}~\frac{1}{x_{0}^{d}}\int d^{d-1}\bar{x}~K(x_{0},\bar{x%
};\bar{x}_{b})K(x_{0},\bar{x};\bar{y}_{b})
\end{equation*}%
where $K$ is the massive bulk to boundary propagator. The IR divergent
contribution to this integral comes from $x_{0}\sim 0$, where we can
substitute one of the propagators by $K(x_{0},\bar{x};\bar{x}_{b})\sim
(x_{0})^{d-1-\Delta }\delta (\bar{x}-\bar{x}_{b})$. The result is%
\begin{equation*}
\delta m^{2}\int dx_{0}~\frac{1}{x_{0}}\left\vert \bar{x}_{b}-\bar{y}%
_{b}\right\vert ^{-2\Delta }
\end{equation*}%
It is important to note that this expression for the perturbed two point
function could be derived explicitly from the expression of the two point
function as an integral over the boundary. One simply uses Green's theorem
and a perturbative analysis of the Klein-Gordon equation. The same statement
would \textit{not} be true in AdS/CFT. In that context, the Euclidean
boundary conditions depend on $\delta m^{2}$, and so the straightforward
perturbative analysis of the path integral misses a term coming from the
perturbation of the boundary conditions. It turns out that the missing term
is sub-leading if the boundary operator is irrelevant, but is the dominant
term if it is marginal or relevant.

By contrast, in the one loop computation with massless fields and
non-derivative coupling, we obtained the IR divergent part%
\begin{equation*}
\int dx_{0}\int dy_{0}~\frac{1}{x_{0}^{d}}\frac{1}{y_{0}^{d}}%
(x_{0}y_{0})^{d-1-\Delta }\left( \frac{x_{0}y_{0}}{\left\vert \bar{x}_{b}-%
\bar{y}_{b}\right\vert ^{2}}\right) ^{\Delta }\left( \ln x_{0}+\ln
y_{0}\right)
\end{equation*}%
The first term after the integration measure comes from the two bulk to
boundary propagators, which we have approximated by their small $x_{0}$
limits. This enabled us to do the two spatial integrals using the $\delta $
functions. The first term in square brackets is the asymptotic form of the
massive bulk propagator, while the second is that of the massless
propagator. We note that if we had instead exchanged a massive field from
the principle series in the loop, or if the massless scalar had derivative
couplings, this last factor would have been a positive power of $x_{0}$ and
all the integrals in the loop diagram would have been convergent. This means
that for a purely massive theory the IR region of coordinate space does not
contribute to the mass renormalization at all\footnote{%
We would get contributions from the region where the two bulk points in the
diagram were close together, corresponding to the usual UV mass
renormalization.}. The value of the mass renormalization following from
exchange of a minimal massless scalar, with soft couplings is thus%
\begin{equation*}
\delta m^{2}\propto \int dx_{0}~\frac{1}{x_{0}}\ln x_{0}\sim \ln ^{2}T\sim
\ln ^{2}\Lambda
\end{equation*}%
The last equality reflects our prejudice that the IR cutoff should be
determined in terms of the c.c., by the requirement of finite entropy.

We note that minimally coupled scalars would generally arise as
Nambu-Goldstone bosons and would be derivatively coupled. Our calculation
shows that one would not expect IR mass divergences in models with NG
bosons. However, we believe that there are indications that gravity has IR
divergence problems comparable to those of minimally coupled massless bosons
with soft couplings. Thus, the divergence we have uncovered reflects our
best guess at the behavior of perturbative quantum gravity in dS space.

\section{Generalization to a Model with Gravity}

The simplest generalization of the calculations we have done is to a model
of gravity interacting with a massive scalar in a dS background. The
Lagrangian is
\begin{equation*}
\mathcal{L}=\sqrt{\left\vert g\right\vert }\left[ M_{P}^{2}R-\left( g^{\mu
\nu }\partial _{\mu }\phi \partial _{\nu }\phi +m^{2}\phi ^{2}\right) \right]
\end{equation*}%
As always in perturbative quantum gravity calculations must be done in a
fixed gauge. We first studied this problem in the gauge for fluctuations
around the dS metric defined by
\begin{equation*}
h_{\mu \nu }={\frac{1}{d}}g_{\mu \nu }h+H_{\mu \nu }
\end{equation*}%
\begin{equation*}
g^{\mu \nu }H_{\mu \nu }=0=D^{\mu }H_{\mu \nu }
\end{equation*}%
$g_{\mu \nu }$ is the background dS metric, and $D^{\mu }$ its Christoffel
connection. In this gauge, the Lagrangian for $h$ is that of a scalar field
with tachyonic mass, while the components of $H_{\mu \nu }$ satisfy a
massive Klein-Gordon equation. One might think that the IR divergences at
one loop arise only from the exchange of $h$\footnote{%
In this gauge, ghosts couple only to gravitons and so there are no ghost
contributions to the one loop boundary two point function of the massive
scalar.}. If this were the case, the calculation would be a simple
generalization of our non-derivative trilinear scalar interaction, with the
massless field replaced by a tachyon.

The result of this computation is disastrous and confusing. The IR
divergence is power law rather than logarithmic (relative to the tree level
calculation). Furthermore the power of $\left\vert \bar{x}_{b}-\bar{y}%
_{b}\right\vert $ differs from the tree level power, so we cannot interpret
the effect as a mass renormalization. If this result were valid one would be
led to the conclusion that the dS/CFT correlation functions simply did not
exist, even in perturbation theory, and the divergence could not be
explained as a divergent mass renormalization.

We gained insight by viewing the transverse gauge as the $\alpha \rightarrow
0$ limit of the one parameter family of gauge fixing Lagrangians
\begin{equation*}
\delta \mathcal{L=}\frac{1}{2\alpha }\left( D^{\mu }H_{\mu \nu }+2b\alpha
\partial _{\nu }h\right) ^{2}
\end{equation*}%
The coefficient $b$ is chosen to cancel the mixing between $H_{\mu \nu }$
and $h$ in the classical Lichnerowicz Lagrangian for fluctuations around dS
space. In this class of gauges, it is easy to see that the tachyonic mass,
as well as the overall normalization of the $h$ propagator, is $\alpha $
dependent. The same is therefore true of the power of $T$ and of $\left\vert
\bar{x}_{b}-\bar{y}_{b}\right\vert $ in the the IR divergent part of the $h$
exchange graph.

Thus, either this contribution is canceled by $H_{\mu\nu}$ exchange, or the
answer is not gauge invariant. Formal arguments using graphical Ward
identities seem to suggest that the boundary two point function is indeed $%
\alpha$ independent. Thus, we expect the power law IR divergences to cancel
at this order. This suggests the possibility that logarithmic divergences,
which come from the behavior of the transverse, traceless part of the
graviton propagator, may not cancel. Gravitational theories would then
exhibit the same sort of IR divergences as our toy model. Of course, we
really need to do a careful computation in order to verify gauge invariance
of the results. We plan to return to this in a future publication. See \cite%
{Henningson:1998cd} and references therein.

\appendix

\section{Comparison with AdS}

In this appendix we record comparisons of our computation of three massive
scalars, with an analogous computation of AdS space. The purpose of this is
to verify that there is no analogue of the divergences we have found, even
when one of the scalars is massless. The essential reason for this
difference is that the bulk AdS propagator is constructed only from
normalizable modes. By contrast, in dS space the Euclidean propagator
contains both solutions of the homogeneous wave equation at large proper
distance.

\subsection{Three Scalar Fields AdS}

For comparison we will describe the case of three massive scalar fields with
cubic interaction in AdS.

As before it is easy to see that in AdS\ the part of $G_{\text{\textit{1-loop%
}}}^{\text{\textit{IR}}}(\bar{x}_{1},\bar{x}_{2})$ that is dependent on $%
\epsilon $ is proportional to%
\begin{equation*}
\epsilon ^{2\Delta _{+}}
\end{equation*}%
In AdS we consider only one type of modes%
\begin{equation*}
\Delta =\Delta _{+}=\frac{1}{2}\left( d-1+\sqrt{%
(d-1)^{2}+4m_{i}^{2}R_{AdS}^{2}}\right) =\frac{1}{2}(d-1)\left( 1+\sqrt{%
(1+\alpha _{i})}\right)
\end{equation*}%
with%
\begin{equation*}
\alpha _{i}=\left( \frac{2m_{i}R_{AdS}}{d-1}\right) ^{2}
\end{equation*}%
so%
\begin{equation*}
\Delta _{+}>0,~\forall ~\alpha _{i}
\end{equation*}%
and $G_{\text{\textit{1-loop}}}(\bar{x}_{1},\bar{x}_{2})$ is IR convergent
for every $\alpha _{i}$ even when $m_{i}$ is zero.

\subsubsection{Anti de Sitter:\ Scalar Propagator}

The two-point function for a scalar field of mass $m$ in AdS$^{d}$ has been
derived for example in \cite{Allen:1985wd}. They find%
\begin{eqnarray}
G(z) &=&rz^{-a}F(a,a-c+1;a-b+1;z^{-1})  \label{2-point fnct.: AdS scalar} \\
r &=&\frac{\Gamma (a)\Gamma (a-c+1)}{\Gamma (a-b+1)\pi ^{\frac{d}{2}}2^{d}}%
R^{2-d}  \notag
\end{eqnarray}%
with $a,~b,~c$ given respectively by (\ref{Delta +}), (\ref{Delta -}), (\ref%
{c}) and where for AdS$^{d}$ we have $R=i\tilde{R},~\tilde{R}\in
\mathbb{R}
$.

The asymptotic expansion $z\rightarrow \infty $ of (\ref{2-point fnct.: AdS
scalar}) is

\begin{equation*}
F(a,a-c+1;a-b+1;z^{-1})\rightarrow 1
\end{equation*}

\begin{equation*}
\lim_{z\rightarrow \infty }G(z)\sim rz^{-\Delta }
\end{equation*}%
with

\begin{equation*}
\Delta =\Delta _{+}=a=\frac{1}{2}\left( d-1+\sqrt{(d-1)^{2}+4m^{2}\tilde{R}%
^{2}}\right)
\end{equation*}

\section{Spinor Green Functions\label{Spinor Two-point Function}}

Here we record the spinor Green Functions needed for the computations and
their asymptotic behavior. For a more exhaustive discussion see for example
\cite{Allen:1985wd}, \cite{Anguelova:2003kf}, \cite{Muck:1999mh}.

\subsection{Spinor Parallel Propagator}

In this section we will derive a differential equation for the spinor
parallel propagator $\Lambda (x^{\prime },x)_{\;\beta }^{\alpha ^{\prime }}$
(\ref{parallel propag.: spinor}) whose action on a spinor is

\begin{equation*}
\psi {^{\prime }}(x^{\prime })^{\alpha ^{\prime }}=\Lambda (x^{\prime
},x)_{\;\beta }^{\alpha ^{\prime }}\psi (x)^{\beta }  \label{def big lambda}
\end{equation*}%
this equation for $\Lambda (x^{\prime },x)_{\;\beta }^{\alpha ^{\prime }}$
will be a fundamental ingredient in the derivation of the spinor Green
function $S(x,x^{\prime })$ .

$\Lambda (x^{\prime },x)$ satisfy the following properties
\begin{subequations}
\begin{align}
n^{\mu }\nabla _{\mu }\Lambda (x,x^{\prime })& =0  \label{property
lambda 1}
\\
\Lambda (x^{\prime },x)& =[\Lambda (x,x^{\prime })]^{-1}
\label{property lambda 2} \\
\Gamma ^{\nu ^{\prime }}(x^{\prime })& =\Lambda (x^{\prime },x)\Gamma ^{\mu
}(x)\Lambda (x,x^{\prime })g_{\mu }^{\nu ^{\prime }}(x^{\prime },x)
\label{property lambda 3}
\end{align}%
(\ref{property lambda 1}) follows from the definition of parallel transport
of a spinor along a curve,\ (\ref{property lambda 2}) derive from the fact
that the $\Lambda (x^{\prime },x)$ form a group and (\ref{property lambda 3}%
)\ indicate how to parallel transport the gamma matrices.

Manipulating the previous equations we obtain
\end{subequations}
\begin{equation}
\nabla _{\mu }\Lambda (x,x^{\prime })=\frac{1}{2}(A+C)\left( \Gamma _{\mu
}\Gamma ^{\nu }n_{\nu }-n_{\mu }\right) \Lambda (x,x^{\prime })
\label{d lambda}
\end{equation}%
and%
\begin{equation*}
\nabla _{\mu ^{\prime }}\Lambda (x,x^{\prime })=-\frac{1}{2}(A+C)\Lambda
(x,x^{\prime })\left( \Gamma _{\mu ^{\prime }}\Gamma ^{\nu ^{\prime }}n_{\nu
^{\prime }}-n_{\mu ^{\prime }}\right)  \label{d lamda prime}
\end{equation*}

\subsection{Bulk Two-Point Function}

The spinor Green $S(x,x^{\prime })$ function is defined by the equation

\begin{equation}
\left[ (\slashD-m)S(x,x^{\prime })\right] _{\;\beta ^{\prime }}^{\alpha }=%
\frac{\delta (x-x^{\prime })}{\sqrt{g(x)}}\delta _{\beta ^{\prime }}^{\alpha
}  \label{definition spinor GF}
\end{equation}%
The most general form for $S(x,x^{\prime })$ is%
\begin{equation}
S(x,x^{\prime })=\left[ \alpha (\mu )+\beta (\mu )n_{\nu }\Gamma ^{\nu }%
\right] \Lambda (x,x^{\prime })  \label{Spinor GF general form}
\end{equation}%
with $\alpha (\mu ),~\beta (\mu )$ functions only of the geodesic distance.

Substituting (\ref{Spinor GF general form}) into (\ref{definition spinor GF}%
) and using (\ref{d lambda})\ we obtain two differential equations for $%
\alpha (\mu )$ and $\ \beta (\mu )$
\begin{align}
\beta ^{\prime }+\frac{1}{2}(d-1)(A-C)\beta -m\alpha & =\frac{\delta
(x-x^{\prime })}{\sqrt{g(x)}}  \label{Eqs for alpha beta 1} \\
\alpha ^{\prime }+\frac{1}{2}(d-1)(A+C)\alpha -m\beta & =0,
\label{Eqs for alpha beta 2}
\end{align}%
Combining (\ref{Eqs for alpha beta 1}) and (\ref{Eqs for alpha beta 2}) we
find the following differential equation for $\alpha (\mu )$%
\begin{equation}
\alpha ^{\prime \prime }+(d-1)A\alpha ^{\prime }-\frac{1}{2}%
(d-1)C(A+C)\alpha -\left[ \frac{(d-1)^{2}}{4R^{2}}+m^{2}\right] \alpha =m%
\frac{\delta (x-x^{\prime })}{\sqrt{g(x)}}  \label{Eq for alpha}
\end{equation}

\subsubsection{De Sitter Space: Massive Spinor}

To derive $S(x,x^{\prime })$ in dS$^{d}$ space we perform the change of
variables%
\begin{eqnarray*}
z &=&\cos ^{2}\frac{\mu }{2R}  \label{change variables alpha z} \\
\alpha (z) &=&\sqrt{z}\gamma (z)  \notag
\end{eqnarray*}%
the Eq. (\ref{Eq for alpha}) become
\begin{subequations}
\label{gammaeq}
\begin{gather}
H(a,b;c;z)\gamma (z)=0  \label{Hypergeom. eq for gamma} \\
H(a,b;c;z)=z(1-z)\frac{d^{2}}{dz^{2}}+[c-(a+b+1)z]\frac{d}{dz}-ab  \notag
\end{gather}%
with
\end{subequations}
\begin{equation*}
a=\frac{d}{2}-i|m|R,\quad b=\frac{d}{2}+i|m|R,\quad c=\frac{d}{2}+1
\label{a, b, c for spinor}
\end{equation*}

As explained in Section \ref{SEC: dS GF},\ the solution corresponding to the
\textit{Euclidean vacuum} is the one that is singular only at $z=1$ i.e.
\begin{equation*}
\gamma (z)=\lambda \,\mathrm{F}(a,b;c;z)=\lambda \,\mathrm{F}%
(d/2-i|m|R,d/2+i|m|R;d/2+1;z)  \label{gamma for S}
\end{equation*}%
\begin{equation*}
\alpha (z)=\lambda \sqrt{z}\,\mathrm{F}(d/2-i|m|R,d/2+i|m|R;d/2+1;z)
\label{alpha for
S}
\end{equation*}%
The constant $\lambda $ is derived by the requirement that (\ref{Spinor GF
general form}) has the same behavior of the flat spinor Green function for $%
R\rightarrow \infty $. We have
\begin{equation*}
\lambda =-m\frac{\Gamma (d/2-i|m|R)\Gamma (d/2+i|m|R)}{\Gamma (d/2+1)\pi
^{d/2}2^{d}}R^{2-d}  \label{constant lambda for S}
\end{equation*}%
Finally $\beta (z)$ is determined by the Eq. (\ref{Eqs for alpha beta 2})
\begin{align}
\beta (z)& =-\frac{1}{m}\left[ \frac{1}{R}\sqrt{z(1-z)}\frac{d}{dz}+\frac{d-1%
}{2R}\sqrt{\frac{1-z}{z}}\right] \alpha (z)  \label{beta for S} \\
& =-\frac{\lambda }{mR}\sqrt{1-z}\left[ z\,\mathrm{F}%
(d/2+1-i|m|R,d/2+1+i|m|R;d/2+2;z)\phantom{\frac{n}2}\right.  \notag \\
& \quad +\left. \frac{d}{2}\,\mathrm{F}(d/2-i|m|R,d/2+i|m|R;d/2+1;z)\right]
\notag
\end{align}
The asymptotic $z\rightarrow -\infty $ expansion for the spinor two-point
function is found to be
\begin{equation}
\lim_{\substack{ x_{0}\rightarrow 0  \\ ~y_{0}\rightarrow 0}}S(x,y)=\left(
\left( C_{+}\frac{-x_{0}y_{0}}{(\overline{x}-\overline{y})^{2}}\right)
^{\Delta _{+}}+C_{-}\left( \frac{-x_{0}y_{0}}{(\overline{x}-\overline{y})^{2}%
}\right) ^{\Delta _{-}}\right) \frac{\bar{\Gamma}\cdot \left( \overline{x}-%
\overline{y}\right) }{\left\vert \overline{x}-\overline{y}\right\vert }
\label{2-point fnct.:
spinor, asympt. exp.}
\end{equation}%
with%
\begin{eqnarray*}
\Delta _{+} &=&\frac{d-1}{2}+im \\
\Delta _{-} &=&\frac{d-1}{2}-im
\end{eqnarray*}

\subsection{Bulk to Boundary Propagators: dS/AdS\label{Sect.: B-B prop.
spinor}}

The complete expression for the spinor Bulk to Boundary propagators:

\begin{equation}
\psi _{1}(x)=\int d^{d-1}\bar{x}~K(x,\bar{x})\psi _{0}(\bar{x})
\label{Bulk to Bndry: spinor psi}
\end{equation}%
\begin{equation}
\bar{\psi}_{1}(x)=\int d^{d-1}\bar{x}~\bar{\psi}_{0}(\bar{x})K(x,\bar{x}%
)\psi _{0}(\bar{x})  \label{Bulk to Bndry: spinor psi bar}
\end{equation}%
has been given for example in \cite{Henningson:1998cd}.

For our purposes we will need only the asymptotic expansion $%
x_{0}\rightarrow 0,~y_{0}\rightarrow 0$ for the propagators (\ref{Bulk to
Bndry: spinor psi}), (\ref{Bulk to Bndry: spinor psi bar}), we have
\begin{equation}
\lim_{x_{0}\rightarrow 0}(x_{0})^{-\frac{d}{2}+m}\left( -\frac{1}{c}\right)
\psi (x)=\psi _{0-}(\bar{x})-\frac{1}{c}\int d^{d-1}\bar{y}~\left\vert \bar{x%
}-\bar{y}\right\vert ^{-d-1+2m}(\bar{x}-\bar{y})\cdot \bar{\Gamma}\psi _{0+}(%
\bar{y})  \label{Bulk to Bndry: spinor psi, asympt. exp.}
\end{equation}%
\begin{equation}
\lim_{x_{0}\rightarrow 0}(x_{0})^{-\frac{d}{2}+m}\left( \frac{1}{c}\right)
\bar{\psi}(x)=\bar{\psi}_{0+}(\bar{x})+\frac{1}{c}\int d^{d-1}\bar{y}~\bar{%
\psi}_{0-}(\bar{y})(\bar{x}-\bar{y})\cdot \bar{\Gamma}\left\vert \bar{x}-%
\bar{y}\right\vert ^{-d-1+2m}
\label{Bulk to Bndry: spinor psi
bar, asympt. exp.}
\end{equation}

where the constant is $c=\pi ^{d/2}\Gamma (m+\frac{1}{2})/\Gamma (m+\frac{d+1%
}{2})$. And we have used the following decomposition for the fields

\begin{eqnarray*}
\psi _{0}(\bar{x}) &=&\psi _{0+}(\bar{x})+\psi _{0-}(\bar{x}) \\
\bar{\psi}_{0}(\bar{x}) &=&\bar{\psi}_{0+}(\bar{x})+\bar{\psi}_{0-}(\bar{x})
\end{eqnarray*}%
with%
\begin{eqnarray*}
\Gamma ^{0}\psi _{\pm }(\bar{x}) &=&\pm \psi _{\pm }(\bar{x}) \\
\bar{\psi}_{\pm }(\bar{x})\Gamma ^{0} &=&\pm \bar{\psi}_{\pm }(\bar{x})
\end{eqnarray*}%
For the right-hand side of (\ref{Bulk to Bndry: spinor psi, asympt. exp.}), (%
\ref{Bulk to Bndry: spinor psi bar, asympt. exp.}) to be integrable, with
respect to the measure $d^{d-1}\bar{y}$ on the boundary $\Sigma $ we have to
impose the conditions%
\begin{eqnarray*}
\psi _{+}(\bar{y}) &=&0 \\
\bar{\psi}_{-}(\bar{y}) &=&0
\end{eqnarray*}


\acknowledgments

TB and LM would like to acknowledge conversations with J. Maldacena about
his approach to the dS/CFT correspondence. Their work was supported in part
by the DOE under grant DE-FG03-92ER40689. The work of WF was supported in
part by the NSF under grant 0071512.

\bibliographystyle{utcaps}
\bibliography{acompat,gravitino_mass_bibtex}

\end{document}